\documentclass[twocolumn]{aastex631}
\usepackage{amsfonts}       
\usepackage{graphicx}
\usepackage{amsmath}
\usepackage{wrapfig}
\usepackage{gensymb}

\makeatletter
\DeclareRobustCommand{\hi}{%
  \mbox{H\check@mathfonts\fontsize\sf@size\z@\selectfont I }%
}
\makeatother

\makeatletter
\DeclareRobustCommand{\hii}{%
  \mbox{H\check@mathfonts\fontsize\sf@size\z@\selectfont II }%
}
\makeatother

\makeatletter
\DeclareRobustCommand{\cii}{%
  [\mbox{C\check@mathfonts\fontsize\sf@size\z@\selectfont II}] %
}
\makeatother

\def \edit#1{\textbf{\color{blue}#1}}

\shorttitle{Low-Frequency RRLs}
\shortauthors{Vydula et al.}

\begin{document}

\title{Low-Frequency Radio Recombination Lines Away From the Inner Galactic Plane}

\correspondingauthor{Akshatha K. Vydula}
\email{vydula@asu.edu}
\author[0000-0002-6611-2668]{Akshatha K. Vydula}
\affiliation{School of Earth and Space exploration,Arizona State University,Tempe, AZ 85281, USA}

\author[0000-0002-8475-2036]{Judd D. Bowman}
\affiliation{School of Earth and Space exploration,Arizona State University,Tempe, AZ 85281, USA}

\author{David Lewis}
\affiliation{School of Earth and Space exploration,Arizona State University,Tempe, AZ 85281, USA}

\author{Kelsie Crawford}
\affiliation{School of Earth and Space exploration,Arizona State University,Tempe, AZ 85281, USA}

\author[0000-0002-2950-2974]{Matthew Kolopanis}
\affiliation{School of Earth and Space exploration,Arizona State University,Tempe, AZ 85281, USA}

\author[0000-0003-1941-7458]{Alan E. E. Rogers}
\affiliation{Haystack Observatory, Massachusetts Institute of Technology,Westford, Massachusetts 01886, USA}

\author[0000-0003-3059-3823]{Steven G. Murray}
\affiliation{School of Earth and Space exploration,Arizona State University,Tempe, AZ 85281, USA}

\author[0000-0003-2560-8023]{Nivedita Mahesh}
\affiliation{Department of Astronomy, California Institute of Technology, Pasadena, CA 91125, USA}

\author[0000-0002-3287-2327]{Raul A. Monsalve}
\affiliation{Space Sciences Laboratory, University of California Berkeley, CA 94720, USA}
\affiliation{School of Earth and Space exploration,Arizona State University,Tempe, AZ 85281, USA}
\affiliation{Facultad de Ingeniería, Universidad Católica de la Santísima Concepción, Alonso de Ribera 2850, Concepción, Chile}

\author[0000-0002-2871-0413]{Peter Sims}
\affiliation{Department of Physics and McGill Space Institute, McGill University, Montreal, QC H3A 2T8, Canada}

\author[0000-0001-5298-1478]{Titu Samson}
\affiliation{School of Earth and Space exploration,Arizona State University,Tempe, AZ 85281, USA}



\begin{abstract}

Diffuse radio recombination lines (RRLs) in the Galaxy are possible foregrounds for redshifted 21~cm experiments. We use EDGES drift scans centered at $-26.7\degree$~declination to characterize diffuse RRLs across the southern sky. We find RRLs averaged over the large antenna beam ($72\degree \times 110\degree$) reach minimum amplitudes between right ascensions~2-6~h.   In this region, the C$\alpha$ absorption amplitude is $33\pm11$~mK (1$\sigma$) averaged over 50-87~MHz ($27\gtrsim z \gtrsim15$ for the 21~cm line) and increases strongly as frequency decreases. C$\beta$ and H$\alpha$ lines are consistent with no detection with amplitudes of $13\pm14$ and $12\pm10$~mK (1$\sigma$), respectively.  At 108-124.5~MHz ($z\approx11$) in the same region, we find no evidence for carbon or hydrogen lines at the noise level of 3.4~mK (1$\sigma$). Conservatively assuming observed lines come broadly from the diffuse interstellar medium, as opposed to a few compact regions, these amplitudes provide upper limits on the intrinsic diffuse lines.  The observations support expectations that Galactic RRLs can be neglected as significant foregrounds for a large region of sky until redshifted 21~cm experiments, particularly those targeting Cosmic Dawn, move beyond the detection phase.  We fit models of the spectral depedence of the lines averaged over the large beam of EDGES, which may contain multiple line sources with possible line blending, and find that including degrees of freedom for expected smooth, frequency-depedent deviations from local thermodynamic equilibrium (LTE) is preferred over simple LTE assumptions for C$\alpha$ and H$\alpha$ lines.  For C$\alpha$ we estimate departure coefficients $0.79<b_n\beta_n<4.5$ along the inner Galactic Plane and $0<b_n\beta_n<2.3$ away from the inner Galactic Plane.


\end{abstract}

\keywords{Radio spectroscopy (1359), Milky Way Galaxy (1054), Interstellar medium (847), Reionization (1383)}

\section{Introduction}

Radio recombination lines (RRLs) have long been a valuable tool used to study the interstellar medium in our Galaxy. These lines appear in the spectra of \hi and \hii regions along the plane of the Galaxy \citep{lockman1996detection}, and act as a probe of the physical conditions in the regions where they are observed. 
Line width and integrated optical depth give information about the temperature, electron density, and emission measure of the gas \citep{salgado2017low, salas2017lofar}. \cite{oonk2017carbon} show that low-frequency carbon and hydrogen lines from molecular clouds can give a lower limit to the cosmic-ray ionization rate. A quantitative study of the low-frequency recombination lines and how they can be used to distinguish hot and cold components of the interstellar medium is done by \cite{shaver1975characteristics, shaver1975theoretical}.

Diffuse recombination lines not associated with any specific star-forming region are also seen along the Galactic Plane and are typically about 0.01-0.1\% of the continuum \citep{erickson1995low}. These lines are usually associated with the colder, lower-density areas inside molecular clouds illuminated by radio-bright objects near the observer's line of sight. Low-frequency carbon line regions are also suggested to be present with photo-dissociation regions \citep{kantharia2001carbon}, stimulated emissions from the low density \hii regions \citep{pedlar1978studies}, \hi self-absorbing regions \citep{roshi2011carbon}, CO-dark surface layers of molecular clouds \citep{oonk2017carbon}, and denser regions within CO emitting clouds \citep{roshi2022arecibo}. Although helium lines have been observed at 1.4~GHz toward the Galactic center \citep{heiles1996radio} and at 750~MHz toward DR21 \citep{roshi2022arecibo}, due to the lower ionization levels, carbon and hydrogen RRLs are brighter and widely used observables to study the interstellar medium at lower frequencies \citep{lowfreqrrl2002}.  \citet{erickson1995low} used the Parkes 64~m telescope with a beam of $\approx$~4$^\circ$ to survey the inner region of the Galactic Plane for carbon RRLs at 76.4~MHz and found a large line-forming region spanning longitudes $|\ell|<20\degree$ and latitudes within a few degrees of the plane. They also observed numerous targets in the plane of the Galaxy in the frequency range of 44~to 92~MHz. C$\alpha$ lines had widths ranging from 5~to 47~km/s.  The region was estimated at distances up to 4~kpc, placing it in the Sagittarius and/or Scutum arms of the Galaxy. They also found lines tangent to the Scutum arm at longitudes between $-48\degree<\ell<20\degree$ which decreased sharply at $\ell=20^\circ$, believed to be due to change in opacity of the absorbing region. \cite{erickson1995low} suggest that the likely sites for the lines are cold \hi regions, however, there were no hydrogen RRL detections.  

One of the highly-studied regions is around the supernova remnant Cassiopeia A (Cas A). It enables observation of both emission and absorption carbon RRLs \citep{payne1989stimulated} and is a common target for low-frequency radio telescopes. \cite{oonk2017carbon} used LOFAR to observe Cas~A from 34~to 78~MHz. They found line widths ranging from 5.5~to 18~km/s, with a line width of 6.27$\pm$0.57~km/s for C467$\alpha$. \cite{payne1989stimulated} observed carbon recombination lines in the frequency range 34~to~325~MHz using the Green Bank telescope in the direction of Cas A and suggest the origins of the lines to be neutral HI regions in the interstellar medium, which is further supported by the evidence presented in \cite{roshi1997hydrogen} using Ooty Radio Telescope at 328~MHz. \cite{roshi2002carbon} performed low and high-resolution surveys of the inner Galaxy over longitudes $-28\degree<\ell<89\degree$ using the Ooty Radio Telescope at 327 MHz, finding carbon RRLs in emission, primarily for $-1\degree<\ell<20\degree$. \cite{kantharia1998carbon} detect carbon lines towards Cas A at 34.5~MHz, 332~MHz, 560~MHz, and 770~MHz, and suggest the line forming regions to be associated with cold atomic hydrogen in the interstellar medium using the integrated line-to-continuum ratio. Spatially resolved carbon RRLs have also been mapped towards Cas A and used as tracers of the cold interstellar gas \citep{salas2018mapping}. Carbon RRLs have also been instrumental in evaluating the physical conditions of the line-forming regions around Orion A \citep{salas2019carbon}.

High-frequency recombination lines are commonly found in dense \hii regions associated with star formation, where the energetic photons ionize the surrounding gas, allowing recombination to occur. This results in RRLs associated with hydrogen, helium, and carbon.  Planetary nebulae are another source, as the expanding ionization bubble around protostars and stellar nurseries provide targets for detection. These types of recombination lines have been observed typically above 1~GHz.  Using data from the \hi Parkes All-Sky Survey at 1.4~GHz to observe H168$\alpha$, H167$\alpha$, and H166$\alpha$ RRLs,  \cite{alves2015hipass} mapped the diffuse lines of the inner Galactic Plane ($-164\degree<\ell<52\degree$) 
and compared the spatial distribution of the ionized gas with that of carbon monoxide (CO).  They reported the first detection of RRLs in the southern ionized lobe of the Galactic Center and found helium RRLs in \hii regions, as well as diffuse carbon RRLs. A higher resolution survey is presented by SIGGMA \citep{liu2019survey} with a beam size of 3.4$'$ using Arecibo L-band Feed Array in the frequency range of 1225~to 1525~MHz spanning H163$\alpha$ to H174$\alpha$. The project THOR uses the Very Large Array (VLA) to survey HI, OH, and recombination lines from the Milky Way at a resolution of 20$"$ in the frequency range of 1~GHz to 2~GHz \citep{bihr2015thor, wang2020hi}.

Beyond our Galaxy, extragalactic RRLs have been seen in both emission and absorption. \cite{shaver1978extragalactic} used the Westerbork Synthesis Radio Telescope to observe H$\alpha$ recombination lines in emission from M82. The Expanded Very Large Array (EVLA) detected hydrogen RRLs in emission from NGC253 \citep{kepley2011unveiling}. More recently, \cite{emig2019first} detected RRLs in the frequency range 109-189.84~MHz in the spectrum of 3C190 with a FWHM of $31.2\pm8.3$~km/s and at a redshift of $z=1.124$. 

With the advent of experiments aiming to detect redshifted 21~cm signals from neutral gas in the intergalactic medium (IGM) between early galaxies at $z>6$, RRLs from our Galaxy and others have been considered as possible foregrounds for the cosmological observations \citep{peng2003foregrounds}. Foregrounds for redshifted 21~cm observations are dominated by Galactic synchrotron radiation, which is typically $\sim$200~K away from the inner Galactic Plane at 150~MHz \citep{{liu2020data, mozdzen2016improved, monsalve2021absolute}} increasing to $\sim$2000~K at 75~MHz \citep{ mozdzen2019low}, and a factor of ten higher along the inner Plane.  These levels are $10^5-10^6$ times larger than the expected 21~cm emission amplitude of 1-10~mK during reionization.  Before reionization, some astrophysical scenarios predict 21~cm absorption of $\sim$100~mK \citep{furlanetto2006cosmology} and non-standard physics models can yield 21~cm signals up to 1000~mK \citep{fialkov2018constraining}. There are strategies in place to mitigate the spectrally-smooth foregrounds from observations, including subtraction based on sky models or parameterized fits and avoidance in either Fourier or delay space (e.g. \citealt{ di2002radio, bowman2009foreground, bernardi2011subtraction, parsons2012per, 2014PhRvD..90b3018L, 2015ApJ...804...14T, 2016MNRAS.458.2928C, 2018ApJ...864..131K, 2019MNRAS.488.2904S}).  

Redshifted 21~cm observations aim to study the early epochs of the Universe $6<z<200$ and target the frequency range of 10-200~MHz. The current generation of instruments primarily aims to make statistical measurements in the Epoch of Reionization ($6<z<13$ or roughly 100-200~MHz \citep{fan2006observational}), where RRLs are typically 0.01-0.1\% of the continuum brightness along the Galactic Plane \citep{erickson1995low, roshi1997hydrogen}. Below 200~MHz, RRLs are typically $\sim$10~K along the Plane.  This is much weaker than the synchrotron foreground but is still larger than the expected cosmological 21~cm signal. RRLs within this frequency range are therefore important for not only characterizing the diffuse gas regions but also for their effects on 21~cm observations. Hydrogen RRLs are primarily expected to be observed in emission, while carbon RRLs are expected to transition from emission above $\sim150$~MHz to absorption below \citep{payne1989stimulated, erickson1995low}.   Away from the Galactic Center and at high-Galactic latitudes where 21~cm observations are generally targeted, RRLs remain poorly quantified \citep{roshi2000hydrogen}.  If diffuse RRLs away from the Galactic Center are about 0.1\% of the synchrotron continuum, similar to their ratio near the Center, we might expect amplitudes as large as 200~mK at 150~MHz increasing to 2~K at 75~MHz.      

A number of experiments are underway to detect and characterize the redshifted 21~cm signal.  They are divided into two groups by observational strategy.  The first group attempts to measure the all-sky averaged global signal and includes EDGES \citep{bowman2018absorption}, SARAS \citep{singh2017first}, LEDA \citep{2018MNRAS.478.4193P}, and REACH  \citep{de2019reach}. The second group aims to detect the power spectrum of angular and spectral fluctuations in the signal and includes HERA \citep{deboer2017hydrogen}, LOFAR \citep{gehlot2019first}, OVRO-LWA  \citep{eastwood201921}, and MWA  \citep{tingay2013murchison}.   In the 50-200~MHz frequency band of redshifted 21~cm observations, diffuse Galactic RRLs potentially form a ``picket fence'' with a $\approx$10~kHz-wide line every $\approx$300-1900~kHz.  The narrow RRLs are easier to mitigate in 21~cm experiments than the brighter, spectrally smooth foregrounds.  In global 21~cm observations, the RRL frequencies can be flagged and omitted from signal analysis with little impact on the result (which we will show in Section~\ref{sec:G21_effects}).  However, for 21~cm power spectrum observations, flagging the RRL frequencies may complicate the analysis by correlating spectral Fourier modes used for the power-spectrum estimate.  This could potentially spread synchrotron contamination into otherwise foreground-free parts of the power spectrum, even after applying inpainting techniques that aim to fill in missing frequency channels.  It would be preferable to skip this flagging if possible. 

Here, we use observations from the EDGES low-band and mid-band instruments to characterize diffuse RRLs at low radio frequencies. Primarily an instrument designed to measure the redshifted global 21~cm, EDGES also provides an opportunity to study the diffuse RRLs in frequencies relevant to 21~cm measurements during the era of Cosmic Dawn and Epoch of Reionization. While the EDGES instruments are very sensitive to faint signals in the radio spectrum between 50 and 200~MHz, they have poor angular resolution. Hence EDGES observations provide primarily an average line strength over large regions of the sky.  We use observed line strengths from EDGES to study the effects of RRLs on global and power spectrum 21~cm observations to determine if detected RRLs and upper limit levels will have a significant impact on redshifted 21~cm analyses. 

In Section~\ref{sec:methods} we summarize the observations and our methods.  In Section~\ref{sec:results} we present the results from EDGES low-band between 50-87~MHz, focusing on the inner Galactic Plane in Section~\ref{sec:innerplane} and away from the inner Plane in Section~\ref{sec:3.3}.  We extend the analysis to include 108-124.5~MHz in Section~\ref{sec:3.6}.  In Section~\ref{sec:21cm} we discuss the implications for 21~cm global measurements and 21~cm power spectrum observations, including foreground cleaning.  We conclude in Section~\ref{sec:4}.  

\begin{figure}
  \hskip-0.5cm
  \includegraphics[scale=0.6]{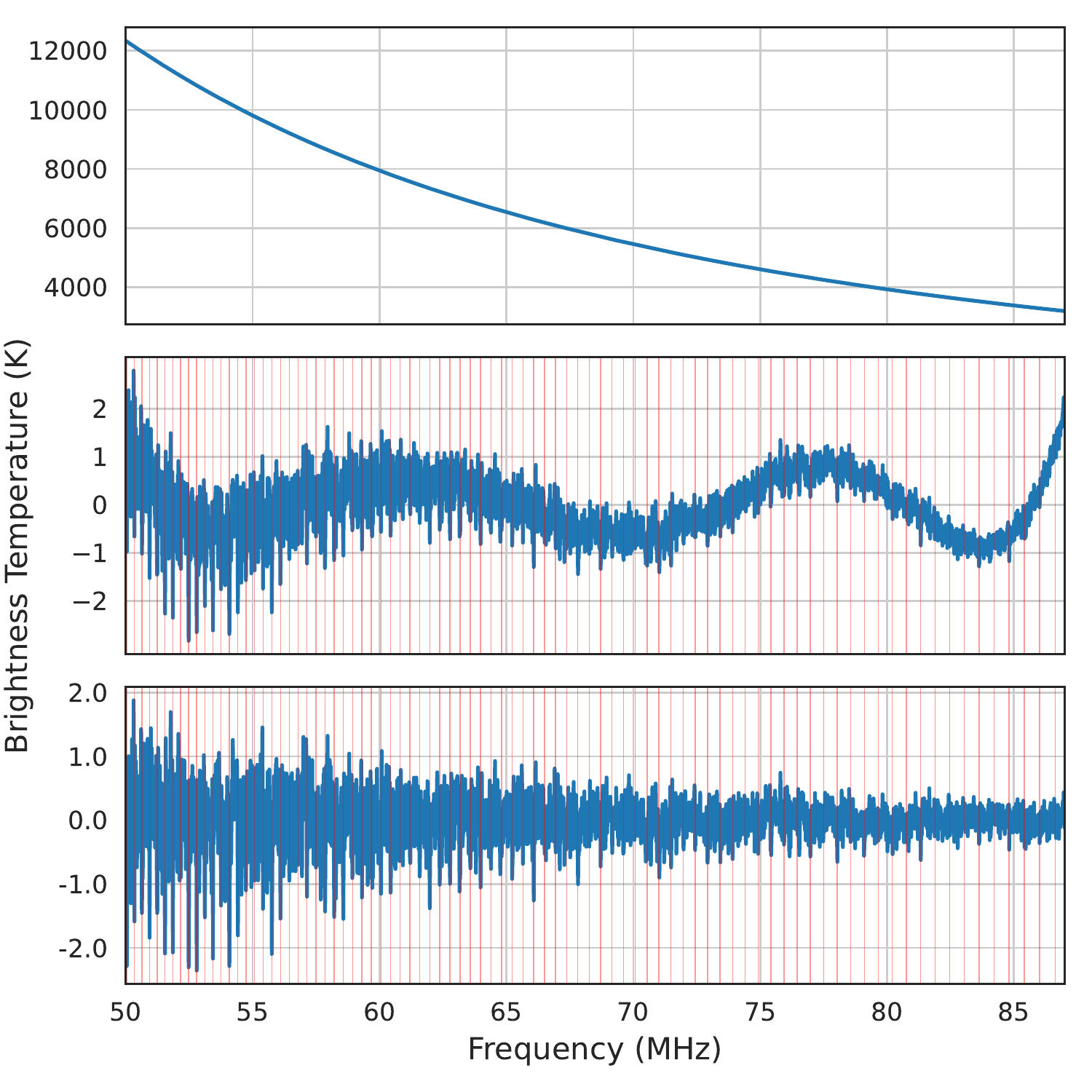}
  \caption{Integrated spectrum (top) and residuals following a 5-term polynomial fit subtraction (middle) and a 9-term polynomial fit subtraction (bottom) for the two-hour bin centered at LST~18~h, when the Galactic Center is at zenith. A 5-term polynomial fit fails to capture all the bandpass features, whereas a 9-term polynomial fit removes all the structures, giving continuum-removed residuals. In the middle and bottom panels, the 85~vertical red lines represent the C$\alpha$ frequencies that are expected within the observed frequency range of 50-87~MHz.  Many of the most extreme negative deviations align with RRL frequencies.  The lines are strongest at low frequencies, reaching about 2~K below 55~MHz, and weaker at higher frequencies with amplitudes of about 1~K at 60~MHz and less than 0.5~K above 70~MHz. }
  \label{fig:specresid}
\end{figure}

\section{Methods}
\label{sec:methods}

The Experiment to Detect the Global EoR Signature (EDGES) is located at the Murchison Radio-astronomy Observatory (MRO) in Western Australia.  It includes multiple instruments, each spanning a subset of the frequency range between 50 and 200~MHz.  Each instrument consists of a wideband dipole-like single polarization antenna made from two rectangular metal plates mounted horizontally above a metal ground plane. Below the ground plane sits the receiver, and signals are carried from the antenna to the receiver via a balun. 

\subsection{Observations}
We use 384~days of observations from the EDGES low-band instrument between October 2015 and April 2017, in the frequency range of 50-100~MHz. The data include and expand on the observations used by \cite{bowman2018absorption} and \cite{mozdzen2019low}.  EDGES is a zenith-pointing drift-scan instrument without any steering capability. This gives the instrument a pointing declination corresponding to the site latitude of -26.7$^\circ$.  At 75~MHz, the antenna beam has a full width half maximum (FWHM) of 72$\degree$ parallel to the excitation axis in the north-south direction and 110$\degree$ perpendicular to the excitation axis \citep{mahesh2021beam}.  The spectrometer samples antenna voltages at 400~MS/s and applies a Blackman-Harris window function and fast Fourier transform to blocks of 65536 samples to yield spectra with 32768 channels from 0-200 MHz. This results in a frequency channel spacing of 6.1~kHz.  Neighboring channels are correlated due to the window function, yielding an effective spectral resolution of 12.2~kHz

The size of the EDGES antenna beam is much larger than any single radio source or region.  
This has two main effects.  First, any one source generally will not contribute substantially to the observed antenna temperature.  Second, many individual sources or regions may be within the beam at any time, especially along the Galactic Plane.  The line strengths and line widths observed by EDGES, therefore,  will be the aggregate effect of many contributions, each with its own intrinsic broadening and Doppler shift.  


The averaging effect of the beam is compounded by the relatively poor spectral resolution of the observations.  At 75~MHz, the raw frequency channel spacing of 12.2~kHz yields an equivalent velocity resolution of 48.8~km/s, which is generally larger than typical gas velocities in RRL regions (see Section~\ref{sec:widths}). 
The end result is that these observations are not suitable for studying individual RRL targets in detail, but rather for characterizing the broad RRL properties in different regions of the sky, particularly away from the Galactic Center.

\begin{figure}
  \centering
  \includegraphics[scale=0.6]{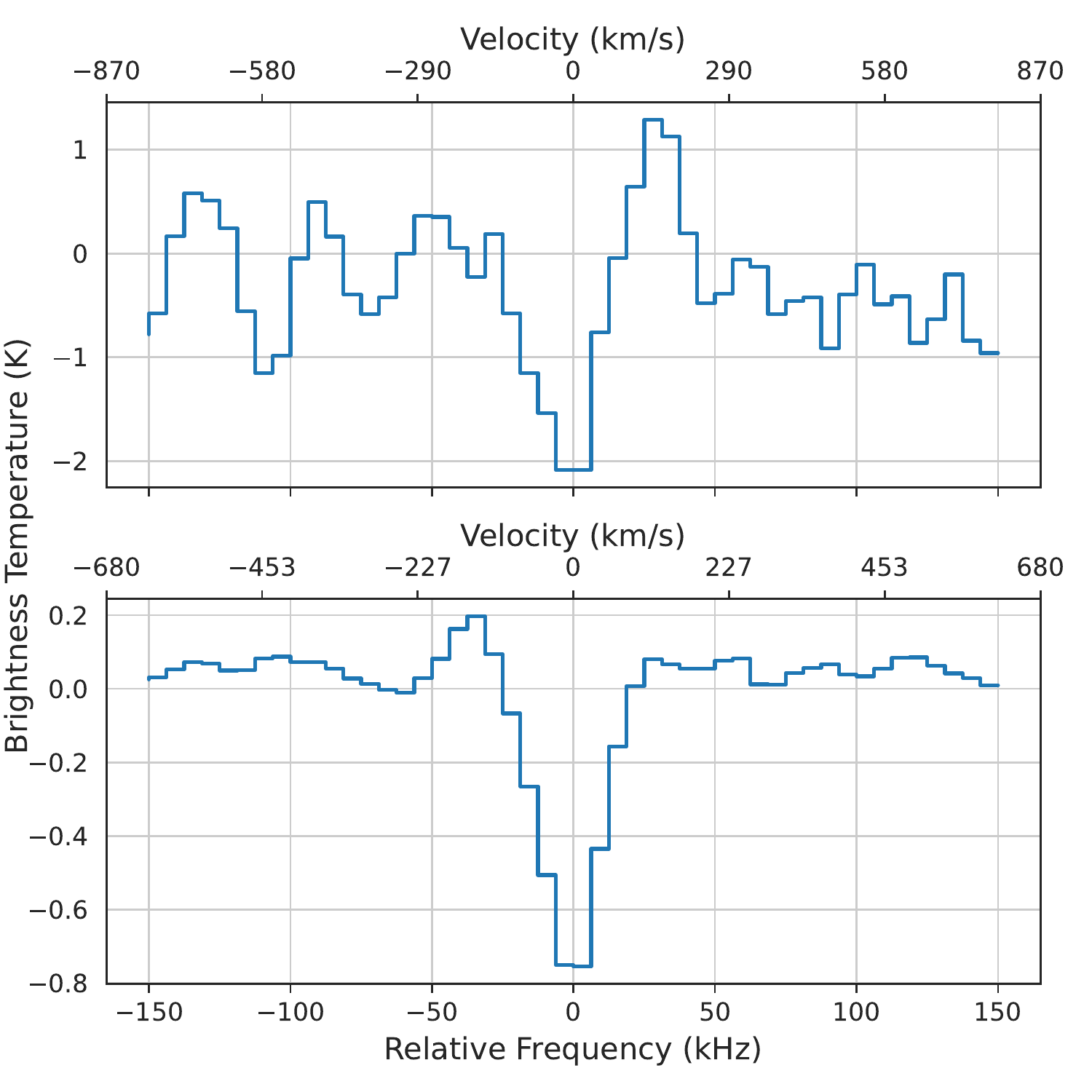}
  \caption{Observed C503$\alpha$ absorption line centered at 51.5~MHz (top) and the stacked average line profile from all 85~RRLs across the band, $423 \leq n \leq 507$ (bottom) for the LST~18~h bin.  The effective frequency of the stacked profile is 66.1~MHz. The significance of detection improves from about 4$\sigma$ for the individual line to 28$\sigma$ for the stacked profile. }
  \label{fig:line_profile}
\end{figure}

\subsection{Data Processing}
\label{sec2.2}

We reduce the EDGES observations and calibrate to an absolute temperature scale following the processes described in \cite{rogers2012calibration}, \cite{monsalve2017calibration}, and \cite{bowman2018absorption}, including filtering of radio-frequency interference (RFI). Following this filter, instances of persistent, weak RFI can still be seen in terrestrial FM radio band transmission above 87~MHz. To avoid complications from this interference, we keep only data below 87~MHz for the remainder of the low-band analysis. The data for each day is then binned into two-hour segments in Local Sidereal Time (LST). Bin centers are referenced to the Galactic Center at LST 17.8~h, yielding a three-dimensional dataset in frequency, day, and LST. For simplicity, we refer to the LST bins by their nearest whole-hour in the rest of this paper (e.g. using LST 18~h to refer to the bin centered on the Galactic Center) since the distinction is insignificant given the two-hour width of each bin and large EDGES beam that spans about 6~h in right ascension.

The spectra are fitted with a 9-term polynomial over the frequency range of 50-87~MHz. The expected RMS in the daily spectra is typically $\sim$5~K when the Galactic Center is overhead and $\sim$2~K when the Galactic Center is below the horizon. We subtract the fits from the polynomial to obtain residuals that have the continuum removed. We perform RFI filtering again, marking any day-LST bins with residuals larger than 30~K (roughly 6$\sigma$) as bad and removing them from the analysis. Typically these cases are due to local weather or abnormal conditions in the upper atmosphere. The resulting total number of spectra for each LST bin varies from 375 to 384.  The residuals for all remaining good days within a given LST bin are then averaged to yield a two-dimensional dataset in frequency and LST.  The duty cycle of the observations was only 17\% of wall-clock time, yielding a total of about 126 hours of effective integration time in each LST bin. Figure \ref{fig:specresid} shows the spectra for LST 18~h and the corresponding residuals after subtracting a polynomial fit, and the expected C$\alpha$ frequencies. The subtraction of the polynomial fit removes the continuum and other broad structures in the spectra. The residual RMS in each spectrum provides an estimate of thermal noise and varies from a high of 374~mK for the spectrum at LST~18~h, when the Galactic Center is overhead, to a low of about 100~mK for LST bins where the plane is primarily near or below the horizon. The noise in each spectrum is strongest at low frequencies because EDGES is sky-noise limited and the sky noise follows the synchrotron power-law spectrum.  

\subsection{Radio Recombination Lines}
\label{sec_rrl}

When an electron in an atom loses or gains energy, it produces an emission or absorption line based on the change in energy levels. An electron falling from a higher energy state, $n_2$, to lower energy state $n_1$ will produce a photon with a frequency \citep[refer][for details]{2002ASSL..282.....G}:
\begin{equation}
    \nu_{n_2 \rightarrow n_1} = \frac{E_{n_2} - E_{n_1}}{h} = c~R~Z^2 \left( \frac{1}{n_{1} ^2}-\frac{1}{n_{2} ^2}\right)
    \label{eq1}
\end{equation}
where $c$ is the speed of light, $R$ is the Rydberg constant, and $Z$ is the effective nuclear charge seen by the electron (usually very nearly $Z=1$ for an excited electron in the outer shell of a neutral atom) for  constants in c.g.s.~units. The Rydberg constant is given by: 
\begin{equation}
R = \frac{2 \pi^2~m_e~e^4}{c~h^3} \left({\frac{M}{M+m_e}} \right)
\end{equation}
where $h$ is Planck's constant, $e$ is the charge of an electron, $M$ is the mass of the nucleus, and $m_e$ is the mass of the electron.  

Free electrons captured by ionized atoms will cascade down through energy levels until they reach the ground state.  Transitions are denoted by the element abbreviation, the principal quantum number of the lower energy level ($n_1$), and a Greek letter indicating the change in principal quantum number ($n_2$ - $n_1$) with $\alpha=1$, $\beta=2$, etc.  
Single ($\alpha$) transitions are the most prevalent and should yield the strongest signals.  Within the EDGES low-band range, neighboring $\alpha$ lines for a given species are generally separated by about 300-600~kHz.   C$\alpha$ and H$\alpha$ lines of the same order are offset from each other by 150~km/s in velocity space, with hydrogen lines lower in frequency due to the lower mass of the hydrogen nucleus compared to carbon.  \citet{erickson1995low} observed that $\beta$ lines are about a factor of two to three weaker than $\alpha$ lines, and $\gamma$ lines are about a factor of four weaker than $\alpha$ lines at similar frequencies, but this ratio is dependent on specific physical conditions of the gas of any line-forming region. $\beta$ and $\gamma$ lines follow similar spacing and offset patterns to $\alpha$ lines in the EDGES band.  


\subsubsection{Line Widths}
\label{sec:widths}

The observed linewidth of a typical RRL is determined by radiation broadening, pressure broadening, and thermal and turbulent Doppler broadening \citep{salas2017lofar, roshi2002carbon}. 
Only Doppler broadening is significant given the EDGES spectral resolution.  The full width half maximum (FWHM) line width due to Doppler broadening is given by:
\begin{equation}
\Delta \nu_G = \frac{\nu_0}{c}  \left [ 4 \ln{2} \left ( \frac {2kT}{M} + V_T^2 \right) \right]^{1/2}
\end{equation}
where $\nu_0$ is the center frequency of the line, $k$ is the Boltzmann constant, $T$ is the temperature of the gas, and $V_T$ is the turbulent velocity (see \citealt{gordon2009radio} for a detailed explanation).  For the most extreme case of hot hydrogen gas at $T=7000$~K \citep{oonk2019spectroscopy}, the characteristic thermal velocity is about 11~km/s.  An example H463$\alpha$ line, near the mean effective frequency of 66.1~MHz for $\alpha$~lines in the EDGES band (see Section~\ref{stacked_profiles_sec}), would have $\Delta \nu_G\approx4$~kHz in the abscence of any turbulent motion.  Thermal broadening is nearly a factor of ten smaller for cold hydrogen gas at $T\approx100$~K.  The thermal velocity of cold carbon gas is only about 0.37~km/s, even lower by an additional factor of 3.5 due to the higher mass of carbon atoms, yielding  $\Delta \nu_G\approx0.1$~kHz for C463$\alpha$.  Turbulent velocities for motions of individual cells of gas are typically $|V_T|\approx20$~km/s for observations on degree scales \citep{2002ASSL..282.....G}.  This is generally larger than the thermal velocities and sets a lower bound of $\Delta \nu_G\approx7$~kHz for hydrogen and carbon line widths in the EDGES band.  

These typical line widths are small compared to the spectral resolution of the EDGES observations. However, the observed lines will include additional Doppler broadening because of other variations in radial velocity between the gas and Earth, including Galactic rotation and Earth's orbit around the Sun.  First, when the Galactic center is at the zenith, the large EDGES beam encompasses the broad molecular ring within about $|\ell|<30^\circ$.  The radial velocity for most of the molecular gas in this region sweeps from about $-60$ to $60$~km/s, increasing linearly with $\ell$
, although some gas in the inner nuclear disk reaches over 200~km/s \citep{dame2001milky}.  Elsewhere along the plane, the radial velocity of molecular gas is generally within $|V_R|<40$~km/s. These Galactic rotational contributions are strong compared to the thermal and turbulent contributions to the line width.  At 66~MHz, this will contribute about 13~kHz to the line width observed by EDGES when the Galactic Center is near the zenith and 7~kHz when other parts of the plane are in the beam.  Second, since the data were recorded over 18~months, the projection toward the gas of Earth's $|V_E|=30$~km/s orbital velocity around the Sun over that period varies.  This further broadens line widths in the long integrations. Regions viewed along the ecliptic plane, including the Galactic Center and anti-center, are broadened by the full 30~km/s, equivalent to nearly 7~kHz.  The effect decreases away from the ecliptic plane.  At a declination of -26.7\degree, the effect is weakest around LST~6~h.  


Treating the radial velocity effects analogously to turbulent motions (but neglecting the FWHM conversion factor since they are not necessarily Gaussian distributions) and summing all of the Doppler velocity broadening contributions, the FWHM line widths can be estimated using:
\begin{equation}
    \Delta\nu_G  \approx  \frac{\nu_0}{c} \left [ 4 \ln 2 \left( \frac {2kT}{M} + V_T^2 \right ) + V_R^2 + V_E^2 \right ]^{1/2}
\end{equation} 
For cold carbon gas, the C463$\alpha$ line near the effective frequency of 66~MHz is expected to have width $\Delta \nu_G\approx16$~kHz towards the Galactic center and 13~kHz away from the Galactic center. For the hot hydrogen case, the H463$\alpha$ line width toward the Galactic center is about 17.5~kHz and 15~kHz away from the Galactic center.  The Doppler broadened line widths are proportional to frequency and will be about 30\% larger at the high end of the observed band.  Even with this additional broadening, individual lines are essentially unresolved by the 12~kHz effective resolution of the observations. 

\begin{figure*}[t]
  \centering
  \includegraphics[scale=0.45]{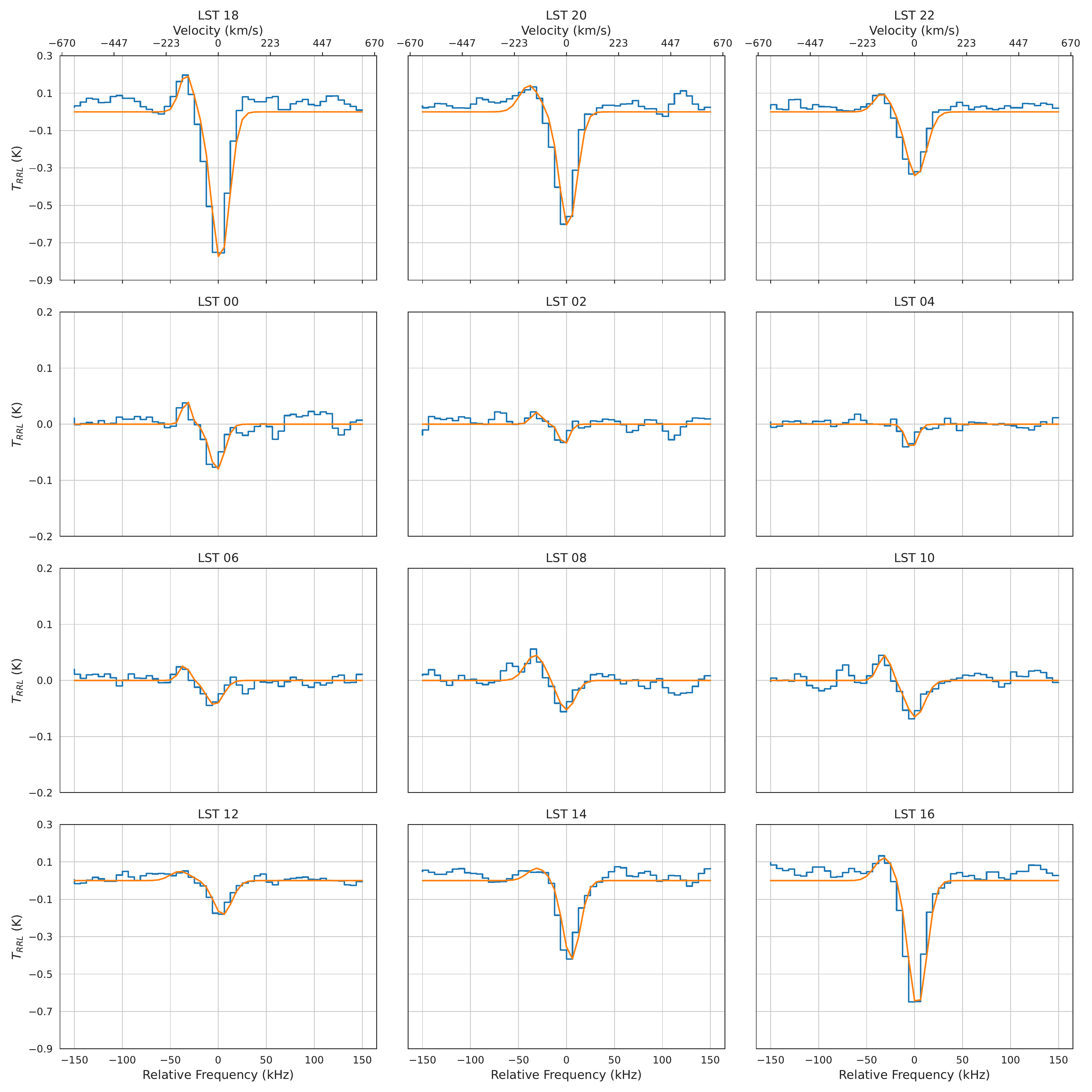}
   \caption{Stacked C$\alpha$ line profile relative to the continuum for each LST bin (blue) and best-fit double Gaussian model (orange).  Each stacked profile is created from 85 individual lines with principal quantum numbers $423 \leq n \leq 507$ and has an effective line center of 66.1~MHz.  The small emission feature about 35~kHz below the C$\alpha$ absorption in each panel is from H$\alpha$. The lines are strong when the Galactic Center is in the beam, peaking at LST~18~h when the Center is at the zenith.  They are weak when higher Galactic latitudes are in the beam (e.g. LST 2-6~h). The best-fit Gaussian profile parameters for each LST bin are listed in Table~\ref{tab:fitparamsalpha}. }
  \label{fig:average_profiles}
\end{figure*}

\begin{figure*}[t]
  \centering
  \includegraphics[scale=0.45]{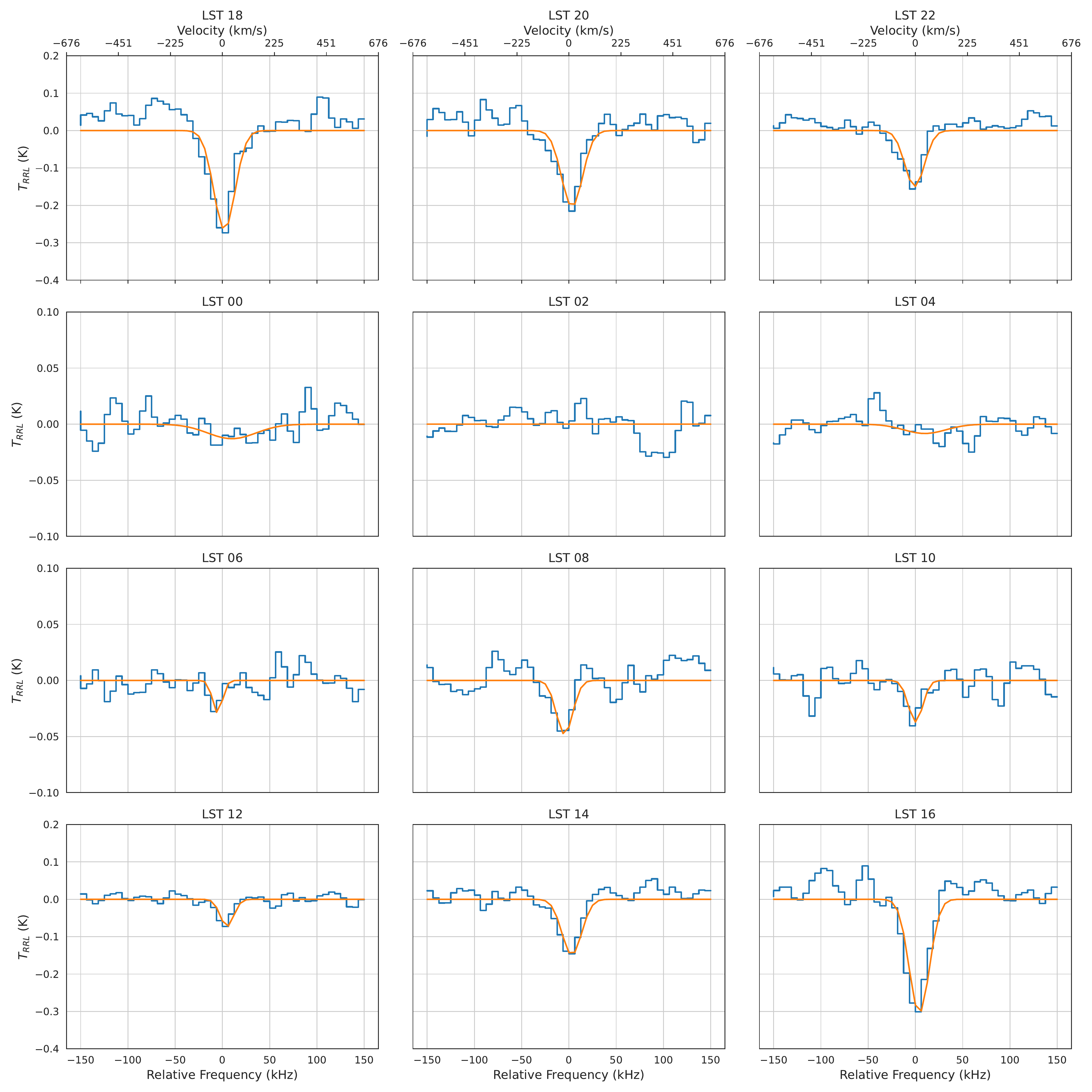}
  \caption{Same as Figure~\ref{fig:average_profiles}, but for $C\beta$ lines.  Each stacked profile is created from 95 individual lines between $533 \leq n \leq 638$ and has an effective line center of 66.3~MHz. The presence of any contribution from hydrogen is less clear here compared to the $\alpha$ lines, hence we have fit only a single Gaussian to the main C$\beta$ profile in each bin.} 
  \label{fig:betasingle}
\end{figure*}

\subsection{Stacked Profiles}
\label{stacked_profiles_sec}
Figure~\ref{fig:specresid} shows the final integrated spectrum from EDGES for the LST~18~h bin when the Galactic Center is at the zenith, as well as the residuals after removing the continuum fit of the spectrum. The RRLs are particularly evident in the residuals below 60~MHz as numerous absorption spikes extending below the noise.  An example of an individual line is shown in Figure~\ref{fig:line_profile}.

The thermal noise in the residual spectrum of each LST bin is sufficiently large that individual RRLs are not detected at high significance.  We, therefore, take one additional step to average all of the individual RRLs together within each LST bin. Using the residual spectrum for each LST bin, we extract a window around each expected carbon RRL, centered on the channel that contains the line center.  This results in a total of 85~windowed spectra for each LST bin with an effective mean frequency of 66.1~MHz.  The effective frequency is below the mid-point of the band because RRLs are more closely spaced at the lower end of the band.  Each extracted window spans 300~kHz and contains 49~frequency channels.  The full-band polynomial fit subtracted earlier in the analyis removed the continuum background for all lines.  The 85~windowed spectra are averaged to yield the stacked profile.  The stacking analysis is performed at the data's true spectral resolution of 6.1~kHz, aligning the frequency channels closest to the line centers, resulting in a maximum misalignment in individual lines of 6.1~kHz.  
Figure  \ref{fig:line_profile} illustrates the improvement in signal-to-noise ratio from a single C$\alpha$ line to the stacked profile for all C$\alpha$ lines in the band.

\section{Results}
\label{sec:results}

The stacked $\alpha$ line profiles for each LST bin are shown in Figure~\ref{fig:average_profiles}.  
In each bin, we see a clear absorption centered on the expected C$\alpha$ line.  In most of the stacked line profiles, we also see a persistent excess in the brightness temperature centered about 35~kHz below the observed absorption line. This emission excess coincides with the expected offset in frequency for the corresponding H$\alpha$ lines for the same principal quantum numbers, as noted in Section~\ref{sec_rrl}. We therefore begin in Section~\ref{gaussian_fit} by simultaneously fitting for C$\alpha$ absorption and H$\alpha$ emission line profiles in the stacked spectrum at each LST bin.  Then we proceed to restack the spectra at C$\beta$ and C$\gamma$ line centers and fit those lines independently. We note that the noise in the stacked profiles shows non-Gaussian noise that is correlated with the Galactic Plane. This could be due to possible narrowband systematics and/or the possible overlapping of some $\alpha$ lines with higher principal quantum number $\beta$ and $\gamma$ lines.  More sophisticated simultaneous fitting techniques could better address this possibility, but is beyond the scope of this work. 
\subsection{Observed Line Profiles}
\label{gaussian_fit}

\begin{table*}[]
\caption{\label{tab:fitparamsalpha} Best-fit Gaussian model parameters and integrated optical depths for each LST bin}

\begin{tabular*}{\textwidth}{ c  c c c c   c c c c   c c c c }
\hline \hline
Bin & \multicolumn{4}{c}{C$\alpha$} & \multicolumn{4}{c}{C$\beta$} & \multicolumn{4}{c}{H$\alpha$} \\
\hline
 LST  & $A$ & $\nu_0$ & FWHM & $\tau d\nu$ & $A$ & $\nu_0$ & FWHM & $\tau d\nu$ & $A$ & $\nu_0$  & FWHM & $\tau d\nu$ \\
(h) & (mK) & (kHz) & (kHz) & (Hz) & (mK) & (kHz) & (kHz) & (Hz) & (mK) & (kHz) & (kHz) & (Hz) \\
\hline


00 &  -81$\pm$10 &   -1$\pm$1 &  18$\pm$3 & 0.3$\pm$0.1 &         -8$\pm$5 &  10$\pm$12 &  64$\pm$23 & 0.2$\pm$0.1 &          44$\pm$15 & -33$\pm$2 &  10$\pm$4 & -0.1$\pm$0.1 \\ 

02 &  -37$\pm$11 &   -2$\pm$2 &  12$\pm$6 & 0.1$\pm$0.1 &           0$\pm$6 &  10$\pm$1 &   47$\pm$44 & 0.0$\pm$0.1 &          21$\pm$10 & -31$\pm$3 &  13$\pm$12 & -0.1$\pm$0.1 \\ 

04 &  -43$\pm$6 &   -3$\pm$1 &  14$\pm$6 & 0.1$\pm$0.1 &          -10$\pm$6 &  10$\pm$15 & 44$\pm$13 & 0.6$\pm$0.4 &           2$\pm$1 & -27$\pm$24 &  24$\pm$57 & -0.0$\pm$0.1 \\ 

06 &  -44$\pm$7 &  -4$\pm$2 &  20$\pm$5 & 0.2$\pm$0.1 &         -32$\pm$10 &  -5$\pm$2 &  13$\pm$2 & 0.1$\pm$0.1 &          27$\pm$8 & -36$\pm$2 &  13$\pm$7 & -0.1$\pm$0.1 \\ 

08 &  -53$\pm$9 &   0$\pm$2 &  21$\pm$5 & 0.2$\pm$0.1 &         -43$\pm$10 &  -5$\pm$2 &  21$\pm$2 & 0.2$\pm$0.1 &          45$\pm$9 & -33$\pm$3 &  24$\pm$7 & -0.2$\pm$0.1 \\ 

10 &  -65$\pm$8 &   1$\pm$1 &  23$\pm$6 & 0.3$\pm$0.1 &         -37$\pm$8 &   0$\pm$2 &  23$\pm$3 & 0.1$\pm$0.1 &          45$\pm$13 & -31$\pm$2 &  16$\pm$5 & -0.1$\pm$0.1 \\ 

12 & -183$\pm$14 &   4$\pm$1 &  22$\pm$2 & 0.8$\pm$0.1 &         -84$\pm$10 &   5$\pm$1 &  16$\pm$5 & 0.2$\pm$0.1 &          50$\pm$27 & -40$\pm$3 &  24$\pm$7 & -0.2$\pm$0.1 \\ 

14 & -420$\pm$28 &   5$\pm$1 &  21$\pm$1 & 1.8$\pm$0.2 &        -162$\pm$21 &   5$\pm$1 &  23$\pm$2 & 0.7$\pm$0.1 &          66$\pm$18 & -31$\pm$5 &  24$\pm$9 & -0.3$\pm$0.2 \\ 

16 & -677$\pm$34 &   3$\pm$1 &  22$\pm$1 & 3.0$\pm$0.2 &        -304$\pm$27 &   4$\pm$1 &  25$\pm$2 & 1.5$\pm$0.2 &         123$\pm$34 & -32$\pm$4 &  24$\pm$7 & -0.6$\pm$0.3 \\ 

18 & -795$\pm$40 &   2$\pm$1 &  22$\pm$1 & 3.4$\pm$0.3 &        -261$\pm$27 &   2$\pm$1 &  29$\pm$2 & 1.3$\pm$0.2 &         203$\pm$46 & -34$\pm$2 &  17$\pm$2 & -0.7$\pm$0.2 \\ 

20 & -616$\pm$33 &   2$\pm$1 &  22$\pm$1 & 2.6$\pm$0.2 &        -211$\pm$29 &   3$\pm$1 &  26$\pm$2 & 1.1$\pm$0.2 &         143$\pm$32 & -39$\pm$3 &  24$\pm$5 & -0.6$\pm$0.2 \\ 

22 & -346$\pm$21 &   2$\pm$1 &  24$\pm$2 & 1.6$\pm$0.2 &        -162$\pm$17 &   -1$\pm$1 &  26$\pm$1 & 0.7$\pm$0.1 &          98$\pm$23 & -34$\pm$1 &  20$\pm$5 & -0.4$\pm$0.2\\

\hline
\end{tabular*}
\tablecomments{Integrated optical depths ($\tau d\nu$) are calculated using Equation~\ref{eqn_taudnu}. Reported C$\alpha$ and H$\alpha$ line profiles have 85~stacked lines $423 \leq n \leq 507$ with an effective mean frequency of 66.1~MHz, while C$\beta$ has 95 stacked lines $533 \leq n \leq 638$ with an effective mean frequency of 66.3~MHz.}
\vspace{2pt}
\end{table*}

The shape of the stacked profile is determined by the line profiles of each of the 85~lines, including their Doppler broadening (about 13-17~kHz), the instrument's spectral resolution (12~kHz), and misalignment of the individual line centers during averaging due to stacking with discrete spectral channels (6~kHz). The first is approximately a Gaussian profile, while the last two are top hat functions. Combining all the broadening contributions results in the effective convolution of a 13-17~kHz distribution with an $12+6=18$~kHz distribution.  For cold carbon near the Galactic Center, the expected stacked line width is about $(18^2 + 16^2)^{1/2} \approx 24$~kHz.  The width is slightly lower when the Galactic plane is out of the beam with $(18^2 + 13^2)^{1/2} \approx 22$~kHz. Similarlly for the hot hydrogen case, the expected stacked line width is about 25~kHz toward the Galactic Center and 23~kHz away from the Galactic Center. This width is similar to the separation between C$\alpha$ and H$\alpha$ lines within the band, which ranges from 25~to 43~kHz with an average of 33.2~kHz. Thus, we expect there to be some line blending between C$\alpha$ and H$\alpha$ lines, motivating simultaneous fits of both lines in the stacked profiles to reduce bias and better reflect the uncertainties on the inferred properties.

In each LST bin, we simultaneously fit Gaussian models for the H$\alpha$ and C$\alpha$ lines to account for any overlap.  The Gaussian model for each line is given by:
\begin{equation}
\phi_{\text{RRL}}(\nu)= A \exp \left[ {- \frac {(\nu-\nu_0)^2}{2\sigma^2}}\right]
\end{equation}
where $A$ is the peak amplitude of the profile, $\nu_0$ is the frequency of the line center, and $\sigma$ is the standard deviation width of the line.  The FWHM of the profile is $2(2 \ln 2)^{1/2}\sigma$.  The thermal noise variance used in the fit of the average line profile is derived from the variances for each window.  A larger 300-channel region is used to derive these variances, with the lines in each window masked out.  The variance is used in the Gaussian fits to estimate the uncertainty on the best-fit parameters.   The maximum hydrogen line width is limited by a prior to 24~kHz to prevent the model from trying to absorb any broad structure in the residuals, particularly when the hydrogen line is weak or not present. 

The best-fit Gaussian profiles for each LST bin are also shown in Figure~\ref{fig:average_profiles}.  Table~\ref{tab:fitparamsalpha} lists the  best-fit parameters.  We find C$\alpha$ line detection significance ranges from 3$\sigma$ at LST~4~h where the best-fit line amplitude is only -33~mK to 28$\sigma$ at LST~18~h where the amplitude is -795~mK.  H$\alpha$ line detections range from no significant detection at LST~4~h to 6$\sigma$ confidence at LST~18~h where the line is in emission with amplitude 203~mK.  

The typical observed line FWHM is about 23~kHz for C$\alpha$ and 20~kHz for H$\alpha$, in reasonable agreement with expectations.  
We also report integrated optical depths in Table~\ref{tab:fitparamsalpha}.  They are calculated as the area under the best-fit Gaussian model for each stacked absorption line profile normalized by the observed continuum sky temperature, $T_{sky}$, at the effective line center frequency.  This is given analytically as:
\begin{equation}
\label{eqn_taudnu}
 \int \tau(\nu) d\nu \approx \frac{\sqrt{2 \pi}~\sigma~|A|}{T_{\text{sky}}(\nu_0)}
\end{equation}
where $\tau(\nu) \equiv |T_{\text{RRL}}(\nu)| / T_{\text{sky}}(\nu)$.  For C$\alpha$, the observed integrated optical depths range from 0.1~kHz at LST~2~h to 3.4~kHz at LST~18~h.  We note the integrated optical depths are affected by beam dilution (see Section~\ref{sec:innerplane} below) if the signal does not fill the full beam.

Figure~\ref{fig:betasingle} shows the stacked profiles centered on C$\beta$ lines in each LST bin. Within the same frequency range of 50-87~MHz, we exlude those lines that overlap with C$\alpha$ within 1~FWHM ($n_1$ = 548, 553, 558, 577, 582, 587, 592, 611, 616, 621, 626) to avoid overlapping of C$\alpha$ and C$\beta$ lines. We stack 95 C$\beta$ lines with principal quantum numbers spanning $533 \leq n \leq 638$, resulting in an effective mean frequency of 66.3~MHz. We find that EDGES is sensitive to these lines with a maximum significance of 13$\sigma$ for a line amplitude of -304~mK at LST~16~h. The C$\beta$ profiles drop to the noise level when the Galactic Center moves out of the instrument beam. Parameters from these fits are included in Table~\ref{tab:fitparamsalpha}. We do not see clear evidence for H$\beta$ lines, which would be centered about 35~kHz below the C$\beta$ lines, but note additional structures in some LST bins (LST~14~h, LST~16~h, LST~18~h, LST~20~h) roughly 50~kHz below the C$\beta$ line centers. We rule out RFI as the cause of these as the structures are correlated with the Galactic Plane, possibly indicating unaccounted for lines or instrumental errors, as discussed above.

\begin{figure}[t]
 \hskip-0.5cm
  \includegraphics[scale=0.6]{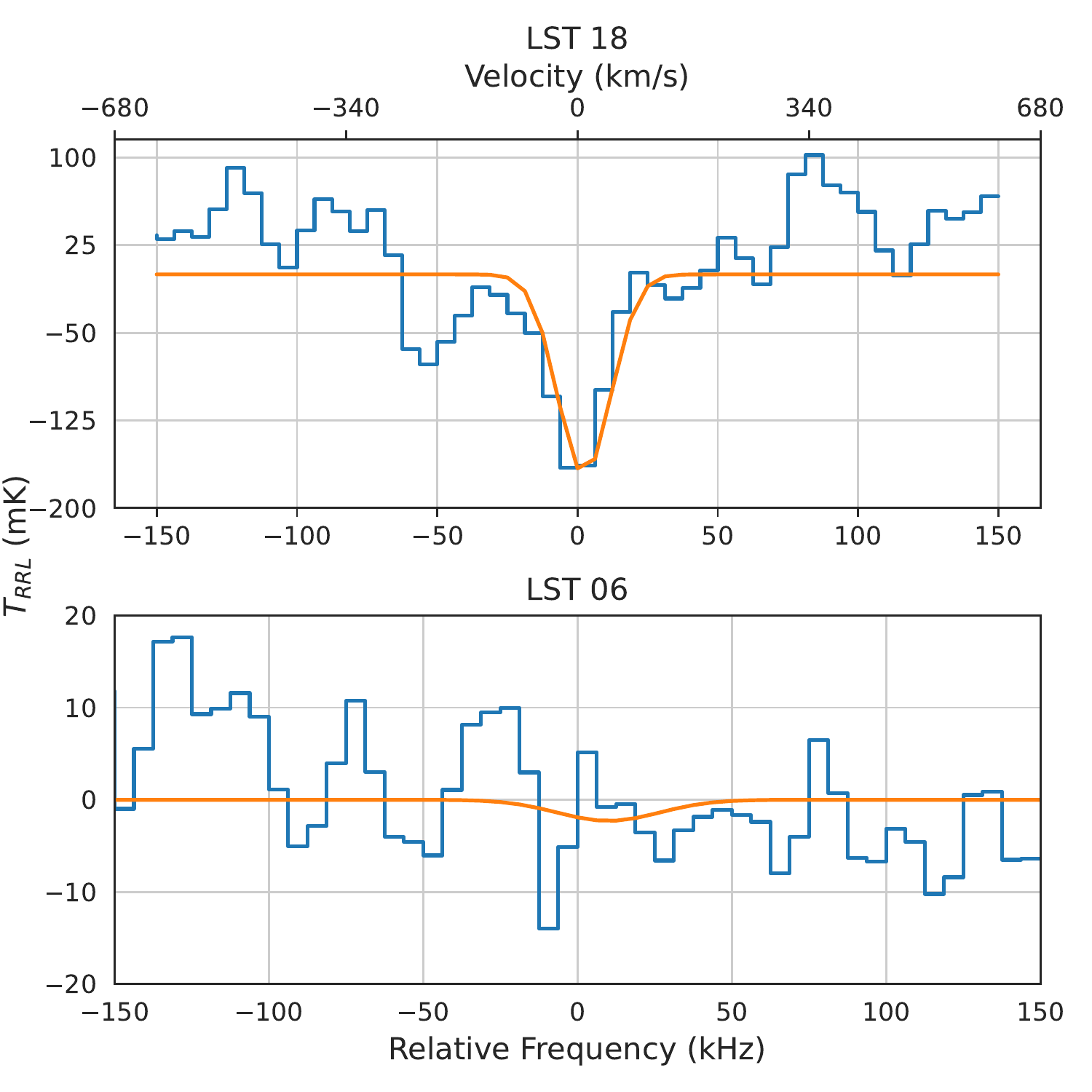}
  \caption{C$\gamma$line profiles, with 122 individual lines stacked, $610 \leq n \leq 731$, and an effective frequency of 66.1~MHz. The Gaussian model in the top panel gives the best-fit C$\gamma$ amplitude of $-171.5\pm34.4$~mK when the Galactic Plane is in the beam at LST~18~h and the bottom panel shows a non-detection when the Galactic Plane moves away from the beam at LST~06~h with a noise level of $\sim$4~mK.}
  \label{fig:gamma_lines}
\end{figure}

Figure~\ref{fig:gamma_lines} shows the stacked profiles at LST~18~h and LST~6~h for C$\gamma$ lines. We stack 122 C$\gamma$ lines with principal quantum numbers spanning $610 \leq n \leq 731$ at an effective mean frequency of 66.1~MHz. The Gaussian fit to the stacked profile in the LST~18~h bin gives a 5$\sigma$ detection of $-171.5\pm34.3$~mK. Similar to C$\alpha$ and C$\beta$ lines, the signal reduces as the Galactic Center moves out of the beam.  We see no evidence for detection between LST~0 and~6~h.

Overall, the magnitudes of C$\beta$ lines are about 30-50\% of the C$\alpha$ lines at corresponding LSTs and C$\gamma$ lines are about 20\% of C$\alpha$ lines, in agreement with trends reported by \cite{erickson1995low}. Table \ref{tab:beamcorrection} gives a summary of all observed line properties at LST~18~h.

\begin{figure*}[t]
  \centering
  \includegraphics[scale=0.5]{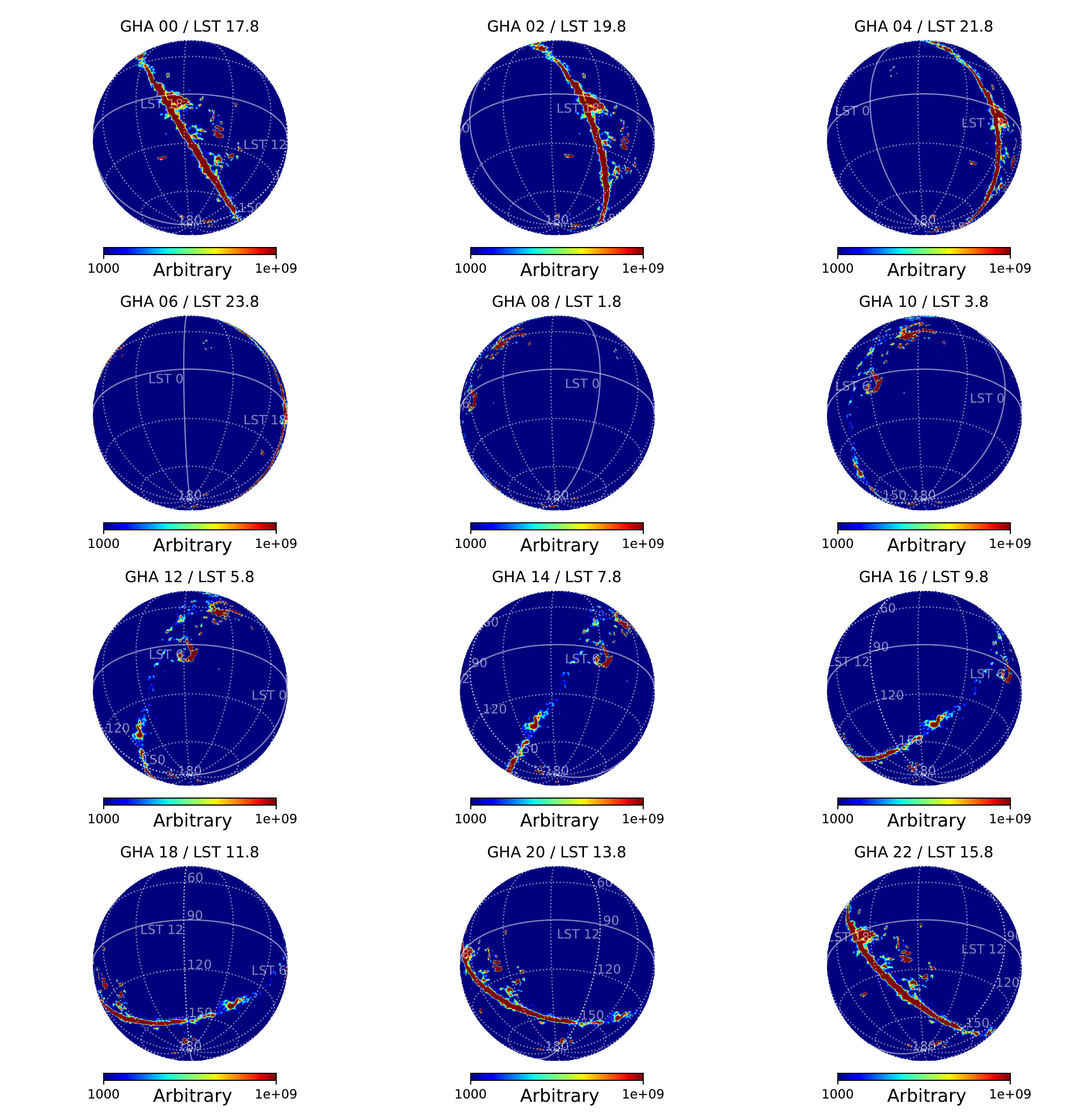}
  \caption{Simulated all-sky RRL proxy maps for the center of each LST bin.  The images are orthographic projections of the Planck~2015 CO map at 115.27~GHz \citep{adam2016planck, ade2016planck}, which provides a rough proxy for the location of carbon in the Galaxy.  The CO line emission is tightly concentrated along the Galactic Plane, similar to RRL surveys of the Plane.}
  \label{fig:fig7}
\end{figure*}

\begin{figure*}[t]
  \centering
  \includegraphics[scale=0.5]{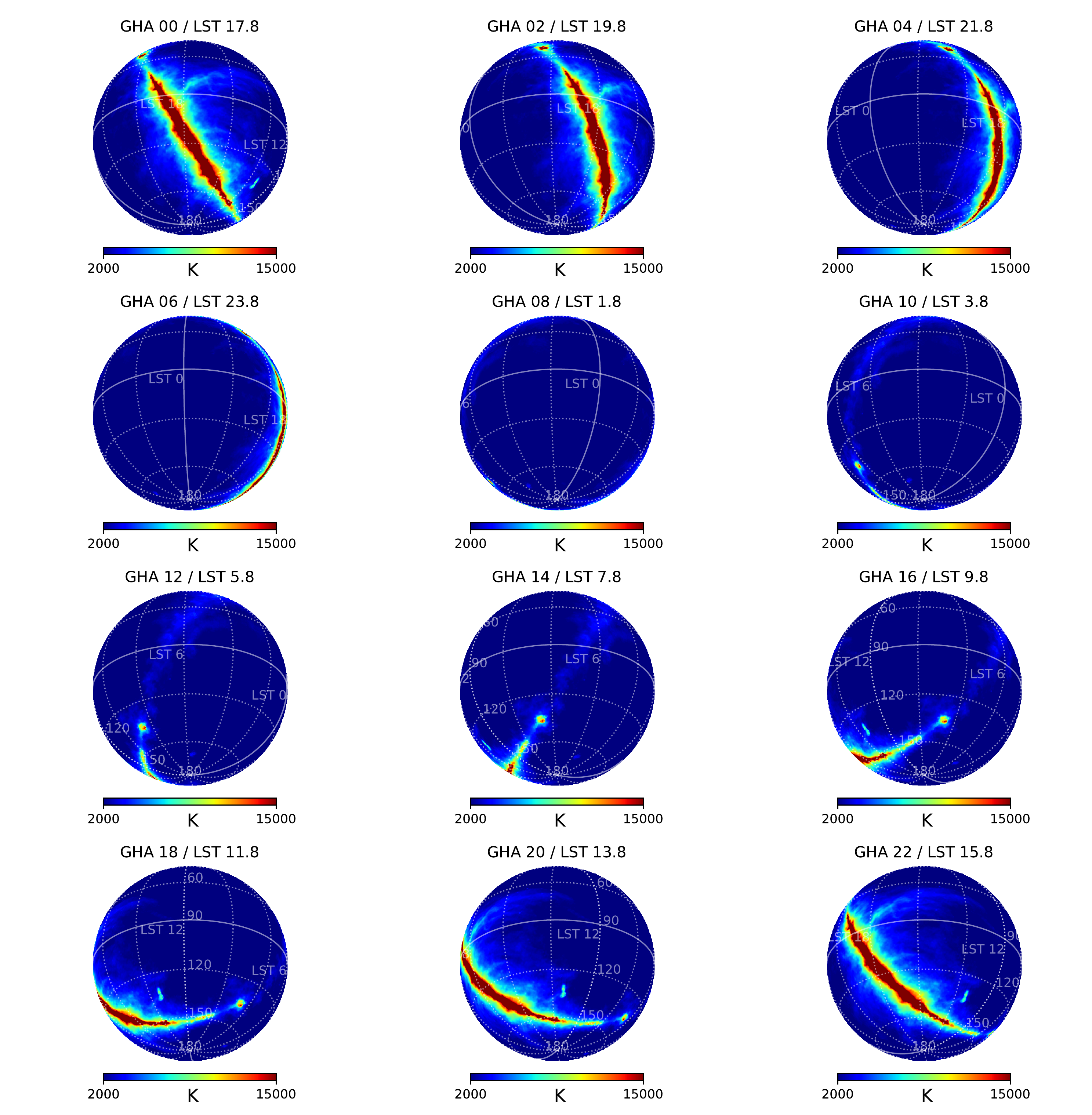}
  \caption{Same as Figure~\ref{fig:fig7}, but for the total radio emission visible by EDGES derived from the \cite{haslam1981408} 408-MHz sky map scaled to 70~MHz using a spectral index of $\beta=-2.4$.  The maps are dominated by synchrotron and free-free emission, which extends from the Galactic Plane to higher latitudes compared to the CO emission observed by Planck.}
  \label{fig:fig6}
\end{figure*}

\subsection{Inner Galactic Plane}
\label{sec:innerplane}

We begin by comparing the EDGES observations to prior measurements along the inner Galactic Plane.  At LST~18~h, when the Galactic Center is directly overhead and the primary response of the EDGES beam includes the strong RRL regions seen in previous surveys, the average line profile shows distinct carbon absorption and hydrogen emission lines.  The C$\alpha$ line has a best-fit Gaussian amplitude of $A=-795$~mK and the H$\alpha$ has a line amplitude of 203~mK.  However, we cannot directly compare the amplitudes observed by EDGES to prior observations with higher angular resolution.  Due to the large size of the EDGES beam, the inner Galactic Plane only fills a relatively small fraction of the total beam. We must account for this beam dilution when the Galactic Plane is overhead to compare EDGES observations with prior measurements.  The large beam is also responsible for spectral effects not present in narrow beam observations.  Narrow beam observations do not span a large range of Galactic longitudes, hence the spectral smearing from Galactic rotation is lower in existing narrow beam measurements.  Lastly, prior measurements have typically been acquired in (or translated to) a single epoch, eliminating the spectral offset due to Earth's orbital velocity that is treated as an additional broadening contribution in our analysis.  Here we will account for these effects in addition to the simple beam dilution.

To calculate the beam dilution, we take the area of the Galactic Plane containing the strong line regions with height $|b|<{3^{\circ}}$ \citep{roshi2002carbon} and longitudinal width $|\ell|<40^\circ$ \citep{alves2015hipass}, yielding an effective solid angle for the Galactic Plane of 480~square degrees.  Comparing this to the effective area of the EDGES beam yields an approximate beam dilution ratio of $f_b=0.088$.  
The effect of spectral smearing in EDGES observations also reduces the peak amplitude of observed line profiles.  
Typical intrinsic widths for low-frequency RRLs along the Galactic Plane are 30~km/s or about 7~kHz at the center of the observed band, compared to 23~kHz observed width for EDGES.  This yields a spectral smearing factor in EDGES observations of $f_s=0.30$.

\begin{table}[h!]
\caption{\label{tab:beamcorrection} Undiluted Line Estimates for LST~18~h}

\hskip-2.0cm
\begin{tabular}{ccccccc}

\hline \hline
Line & n & \multicolumn{2}{c}{Observed} & \multicolumn{3}{c}{Undiluted Estimate}\\
\hline
& & A & FWHM & A & $L/C$ & $\tau d\nu$ \\ 
 & & (K) & (kHz)& (K) & ($10^{-4}$)  & (Hz) \\

\hline
\multicolumn{7}{c}{ 50-87~MHz } \\
\hline
C$\alpha$ & 423-507 & -0.795 & 22 & -30 & -12 & 9 \\
C$\beta$ & 533-638 & -0.261 & 28 & -10 & -4 & 3 \\
C$\gamma$ & 610-731 & -0.171 & 22 & -6 & -2 & 2 \\
H$\alpha$ & 423-507 & 0.203 & 17 & 7 & 3 & -2 \\
\hline
\multicolumn{7}{c}{ 108-124.5~MHz} \\
\hline
C$\alpha$ & 375-392 & -0.046 & 73 & -2 & -2 & 6 \\
H$\alpha$ & 375-392 & 0.098 & 49 & 4 & 5 & -8
\end{tabular}

\tablecomments{Estimates of line strengths that would be observed by a narrow beam instrument for the LST~18~h bin along the inner Galactic Plane.  The effects of dilution due to the large EDGES beam and long observing season are accounted for.}
\end{table}

As an estimate of the undiluted amplitude of the RRL signal from the Galactic Plane, we divide our fitted amplitudes by the calculated beam dilution and spectral smearing ratios for LST~18~h.  Table~\ref{tab:beamcorrection} summarizes the equivalent undiluted line strengths, which broadly match the expectations of $T_{RRL}^{u}\approx10$~K \citep{emig2019first}. As an additional assessment, we report line-to-continuum ratios ($L/C$) and integrated optical depths using the undiluted line amplitudes. The line-to-continuum ratio is calculated using the undiluted peak amplitude of each stacked profile and an estimate of 26,945~K for the undiluted continuum temperature along the inner Galactic Plane at the effective frequency of 66.1~MHz. 
The undiluted integrated optical depth, $\tau d\nu$, is a measure of the total line power. Assuming line blending is minimal (as described in Sec \ref{gaussian_fit}), $\tau d\nu$ is conserved through spectral smearing. Hence, to calculate the undiluted optical depth we apply only the beam dilution factor to the line amplitudes reported in Table~\ref{tab:fitparamsalpha} and use the same estimated undiluted continuum temperature as for the $L/C$ calculations.  We find $10^{-4} \lesssim |L/C| \lesssim 10^{-3}$, and $2\lesssim|\tau d\nu|\lesssim9$.  Both are generally in good agreement with previous measurements.  The undiluted $\tau d\nu$ estimates for C$\alpha$, C$\beta$ and H$\alpha$ match the observations of \citet{oonk2019spectroscopy} within about 20\%.  
We investigate the frequency dependence for each of C$\alpha$, C$\beta$ and H$\alpha$ line strengths further in Section~\ref{sec:rrl_frequency_dependence}.   

\subsection{Away From the Inner Galactic Plane}
\label{sec:3.3}

We are particularly interested in the RRL strength at high-Galactic latitudes and along the outer disk because these are the primary observing regions for redshifted 21~cm experiments.  The best-fit H$\alpha$ and C$\alpha$ amplitudes reduce as the Galactic Plane moves out of the EDGES beam. They reach minima around LST~2-6~h, when the beam is centered around $\ell\approx220\degree$ and $b\approx-50\degree$ and spans a large region of the southern Galactic hermisphere.

The RRLs observed by EDGES away from the inner Galactic Plane may come from a discrete number of small regions, each with strong lines, or they may arise from large areas of diffuse gas with weak lines.  In the scenario of diffuse gas with weak lines, we can assume the gas fills the entire EDGES beam and so requires no dilution factor to estimate its true strength. We use this assumption to place an upper limit on the strength of these signals directly from the observations.  

The inner plane of the Galaxy passes completely out of the EDGES beam starting at about LST~0~h, leaving only high latitudes or the outer disk visible until about LST~10~h.  During this period, C$\alpha$ is always detected at 3$\sigma$ or higher and reaches its weakest amplitude of $-33\pm11$~mK at LST~2~h, as listed in Table~\ref{tab:fitparamsalpha}, when the south Galactic pole is near zenith.  In contrast, the H$\alpha$ line is detected at the beginning and end of the period at 3$\sigma$ or higher, but decreases from about 40~mK at the these boundaries to a minimum of 4$\pm8$~mK at LST~4~h, where it is consistent with no detection.  The 3$\sigma$ upper limit on H$\alpha$ emission at LST~4~h is 28~mK.    Similarly C$\beta$ is marginally detected at about $2\sigma$ at LST~0 and~4~h, but not at LST~2~h.  For the upper limit of C$\beta$ absorption in this region, we use the $3\sigma$ upper limit of the LST~0~h weak detection, giving 32~mK.

\subsection{Model Tracers}

While EDGES observations provide sensitive tests of RRLs away from the inner Galactic Plane, they cannot produce maps of the line strength distribution on the sky. Here we investigate if existing all-sky maps of molecular gas can predict the observed variations in RRL strength with LST, thus potentially serving as proxies for the RRL distribution on the sky.  

We use the Planck 2015 carbon monoxide (CO) map at 115.27~GHz \citep{adam2016planck, ade2016planck}.  CO traces molecular gas in the Galaxy and serves as a proxy for C$\alpha$ strength \citep{chung2019cross,roshi2002carbon}.  We convolve the EDGES beam with the CO map to simulate an observed drift scan as if EDGES were to see the sky described by the map.  For simplicity, we take the CO map as a direct proxy to the RRL line strength.  We do not account for underlying astrophysics and do not attempt to convert CO intensity to CO column density and C column density that could be used to more accurately calculate C$\alpha$ line strengths. Figure~\ref{fig:fig7} illustrates the CO sky as it would be viewed by EDGES.  For comparison,  Figure~\ref{fig:fig6} shows the same view of the total radio emission found by following the same procedure to create simulated observations using the \citet{haslam1981408} 408~MHz map that is dominated by synchrotron radiation.  We assume a constant spectral index of $\beta=-2.4$ for scaling the Haslam map to the observed frequencies.

\begin{figure}[t]
  \centering
  \includegraphics[trim={0.1in 0 0 0},clip, scale=0.6]{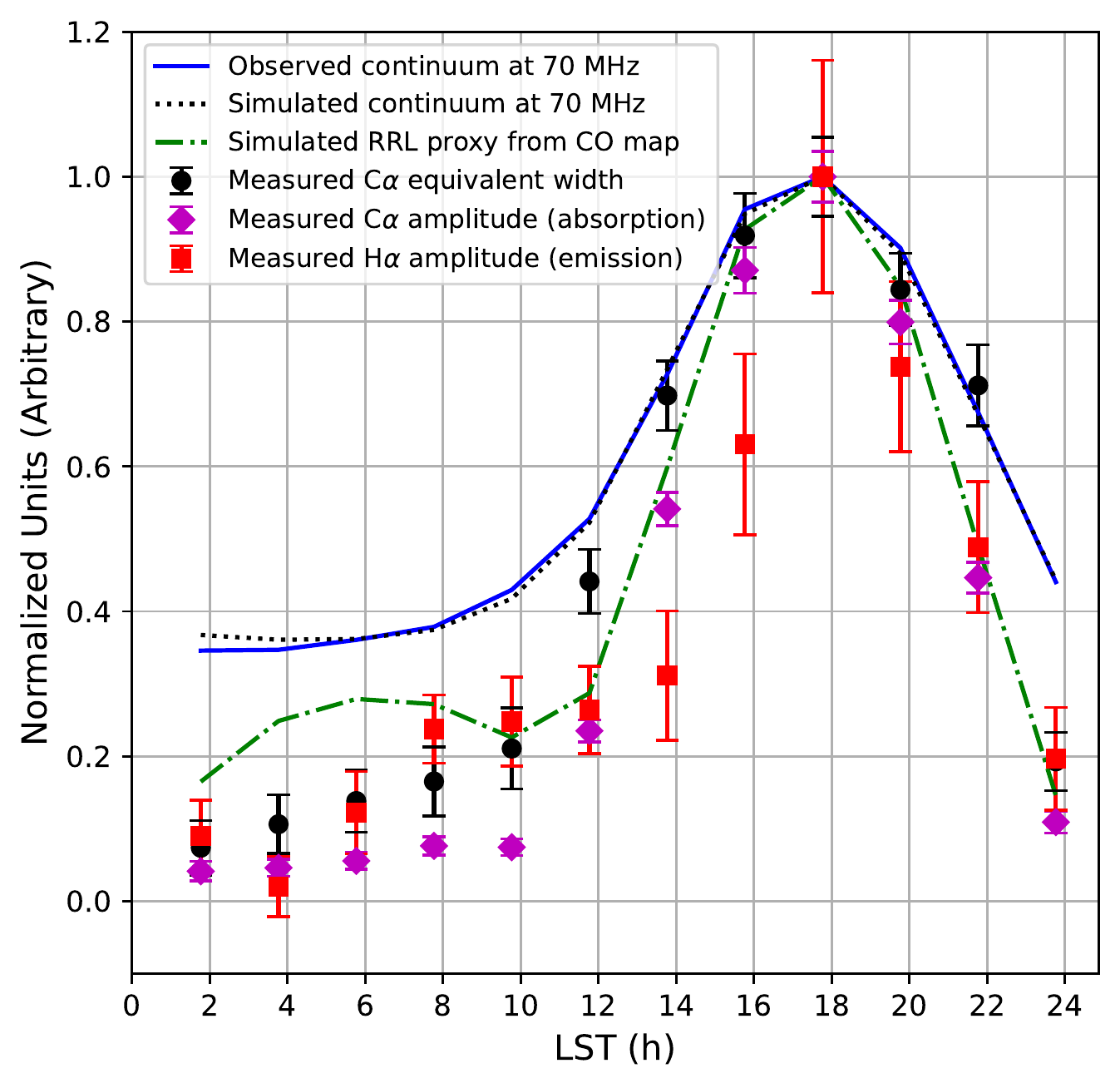}
  \caption{Peak normalized comparison of measured RRL strength and simulated proxy measurements as a function of LST. The average line profile magnitudes for C$\alpha$ (purple diamonds) and H$\alpha$ (red squares) are shown along with the C$\alpha$ integrated optical depths (black circles).  The simulated proxy measurement based on the Planck 115.27~GHz CO~map (dash-dot green line) provides a reasonable predictor of relative RRL strength, particularly when the Galactic Center is in the beam.  However, it over-predicts the RRL strength at low LST.  For reference, the continuum drift scan from EDGES observations (solid blue line) is plotted along with a simulated continuum drift scan using the \cite{haslam1981408} 408-MHz map (dotted black line). As expected, the continuum drift scans show a broader peak than the RRL scans, consistent with synchrotron emission extending farther from the Galactic Plane compared to line emission. }
  \label{fig:fig12}
\end{figure}

In Figure~\ref{fig:fig12}, we show the simulated drift scans along with EDGES C$\alpha$ and H$\alpha$ observations, all normalized to their maximum amplitudes at LST~18~h.  
The CO proxy map reasonably reproduces the relative LST trends of the C$\alpha$ and H$\alpha$ drift scans observed by EDGES, capturing the peak-to-peak range and the temporal width of the cycle more closely than the synchrotron continuum map. This is consistent with RRLs being more concentrated on the Galactic Plane than synchrotron emission. Despite capturing the broad trends, the simplistic CO proxy fails to predict the relative RRL amplitudes and trends with LST observed between LST~0-10~h, suggesting it is a poor template for predicting RRLs in regions of the sky of most interest to redshifted 21~cm power spectrum observations.  

\subsection{Extending to 124.5~MHz}
\label{sec:3.6}
Here we repeat the line analysis for data from the EDGES mid-band instrument using 66~days of observations from early 2020.  The mid-band instrument uses a smaller antenna than the low-band instrument, shifting the frequency range to 60-160~MHz. At 90~MHz, the mid-band antenna beam has FWHM of 75.4$\degree$ parallel to excitation axis in the north-south direction and 106.6$\degree$ perpendicular \citep{monsalve2021absolute}. For this analysis, we reduce the frequency range to 108-124.5~MHz, following a conservative approach of excluding RFI from FM band below 108~MHz and from satellite communication bands and air-traffic control channels above 125~MHz.  Even in this narrow band, some of the frequency channels are dominated by RFI and thus need rigorous flagging. We completely discard time bins that have 60\% or more of the frequency channels flagged. We fit and remove the continuum using an 11-term polynomial and further clean the data by discarding days with residual amplitudes larger than 10~K.  

From Equation~\ref{eq1}, we expect 18~RRL C$\alpha$ transitions ($375 \leq n \leq 392$) in the frequency range of 108~to 124.5~MHz that are roughly spaced 1~MHz apart. We select 600~kHz windows around each line center and stack these segments from the continuum-removed residual spectra resulting in an effective mean frequency of 116.3~MHz. We expect similar LST dependence as low band results and thus, for simplicity, we only investigate two LST bins: 1) a two-hour bin centered on LST~18~h, when the Galactic Center is overhead and 2) a large twelve-hour bin centered on LST~6~h, when the Galactic Center is not overhead. The average line profile was fit in each bin with a double Gaussian model for hydrogen emission and carbon absorption.

\begin{figure}[t]
  \centering
  \includegraphics[scale=0.6]{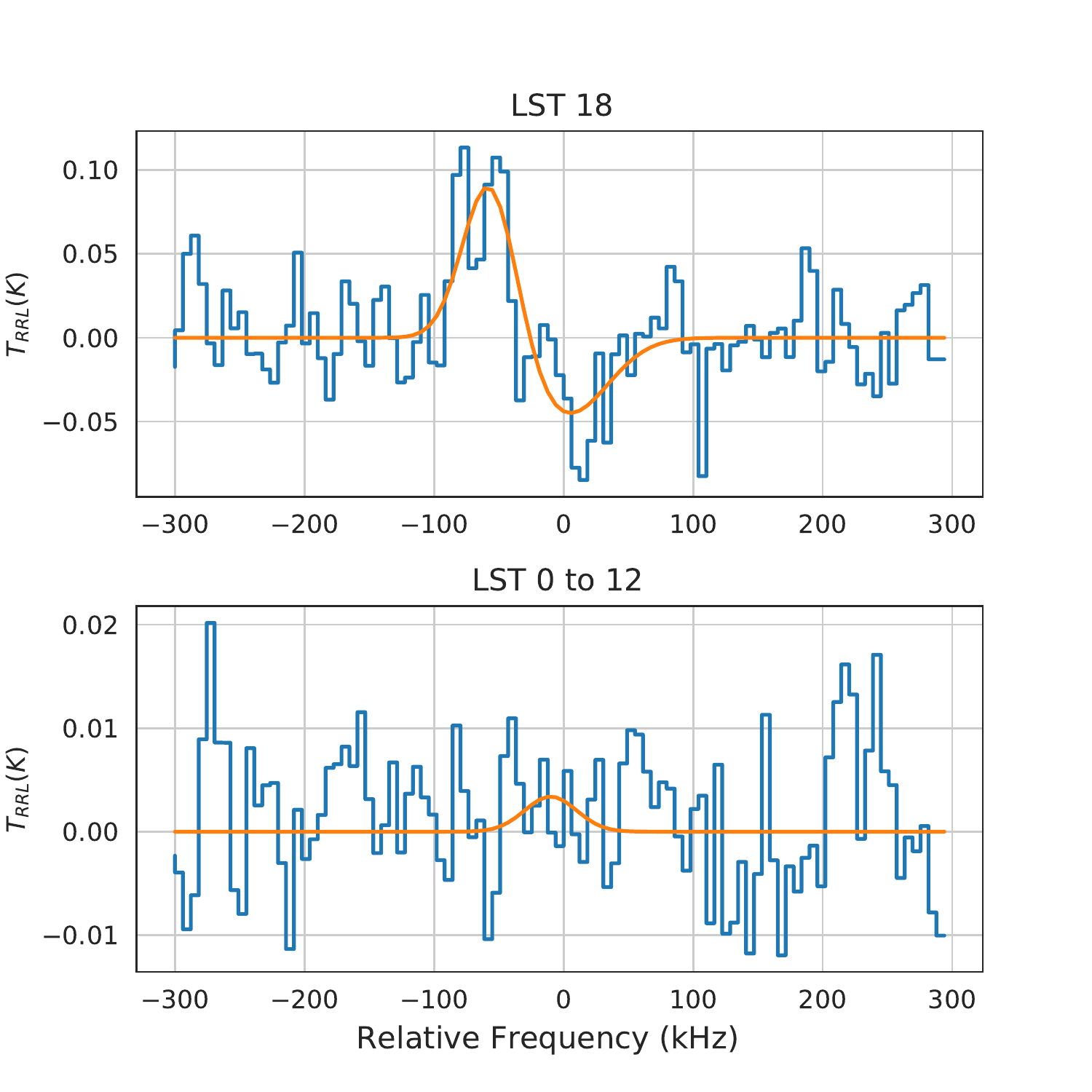}
  \caption{Stacked line profiles (blue) in mid-band EDGES data spanning the frequency range 108-124.5~MHz, with $375 \leq n \leq 392$ and effective mean frequency of 116.3~MHz.  The top panel shows a two-hour LST bin centered on the Galactic Center.   A double Gaussian model (orange) gives a best-fit C$\alpha$ amplitude of $-46 \pm 15$~mK.  The best-fit H$\alpha$ amplitude of $98\pm38$~mK, peaking 59~kHz below the carbon absorption, is about half of the peak level found at lower-frequencies. The bottom panel shows a long twleve-hour bin centered on LST~6~h, yielding stacked line profiles consistent with no detection (best-fit C$\alpha$ amplitude of $3.4\pm3.5$~mK).}
  \label{fig:fig9}
\end{figure}

In the LST 18~h bin, at a center frequency of 116.3~MHz, we find a best-fit amplitude of $-46 \pm 15$~mK and FWHM of $73.4\pm56$~kHz for C$\alpha$ and $98 \pm 38$~mK and FWHM of $48.9\pm15~$kHz for H$\alpha$, as shown in Figure~\ref{fig:fig9}.  We can also correct for beam dilution and compare these observations to existing RRL surveys. The EDGES mid-band antenna is a scaled copy of the low-band antenna, hence it has similar beam properties and the dilution factors we calculated earlier are still valid.  Accounting for the dilution effects and using an undiluted continuum temperature of 6980~K, we find an undiluted H$\alpha$ amplitude of about 4~K with $\tau dv\approx-8$~Hz and $L/C\approx5\times10^{-4}$.  
C$\alpha$ has an undiluted amplitude of 2~K with $\tau d\nu\approx6$~Hz with $L/C\approx-2\times10^{-4}$. Away from the Galactic Center in the large LST bin centered on 6~h, we find the best fit amplitude for C$\alpha$ of $3.4 \pm 3.5$~mK, consistent with no detection, and yielding a $3\sigma$ upper limit of 14~mK.  Similarly, we see no evidence of H$\alpha$ emission present and can assign a comparable upper limit. 

\subsection{Frequency Dependence of RRLs}
\label{sec:rrl_frequency_dependence}

In this section, we investigate the frequency dependence of C$\alpha$, C$\beta$, and H$\alpha$ line amplitudes within the observed bands and compare the EDGES observations to  prior observations and theoretical expectations.  The observed temperature brightness of the lines relative to the background is dependent on the electron temperature ($T_e$) \citep{shaver1980}, optical depth of the lines ($\tau^{LTE}_L$), and background radiation temperature ($T_{R}$).
The assumption of local thermodynamic equilibrium (LTE) may not hold for $n\gtrsim100$ \citep{2017ApJ...837..141S}, resulting in increased contributions from stimulated emission or absorption.  In this regime, observed line temperatures relative to the total continuum can be written as:
\begin{equation}
\label{eqn:rrl_frequency_dependence}
T_{RRL}(\nu_n) \approx \tau^{LTE}_L \left [ b_n T_e - b_n \beta_n T_R(\nu_n) \right ]
\end{equation}
where $b_n$ and $\beta_n$ are departure coefficients \citep{2002ASSL..282.....G} accounting for deviations from LTE and $\nu_n$ is the frequency corresponding to a specific transition line.  For the low-frequency, large-$n$ lines observed here, $\tau^{LTE}_L$ is effectively constant with frequency due to the nearly identical Saha-Boltzmann occupations for nearby $n$ and low photon energy relative to the excitation temperature.  The departure coefficients are smooth functions of $n$, although they may have multiple inflection points.  For C$\alpha$ they can span $0.6\lesssim b_n \lesssim 1.3$ and $-20 \lesssim b_n \beta_n \lesssim 15$, depending on the density and temperature of the gas \citep{2017ApJ...837..141S}, which is expected to be $n_e\approx0.1$~cm$^{-3}$ and $T_e\approx50-100$~K. 

The sky continuum temperature is dominated by synchrotron and free-free emission at low radio frequencies and follows a power-law, $T_{cont}=T_{100} (\nu/\nu_{100})^{\beta}$~K, where $\beta$ is the spectral index and $T_{100}$ is a normalizing temperature at $\nu_{100}=100$~MHz.  The observed sky temperature from Earth toward the inner Galactic Plane is about 10,000~K at 100~MHz, falling to about 1000~K away from the Plane \citep{2017MNRAS.469.4537D}. The gas that produces RRLs is at varying distances from Earth.  Some of the synchrotron and free-emission we observe originates between Earth and the gas.  This portion of the received emission does not experience RRL absorption in the gas and is not included in the radiation background, $T_R$, in Equation~\ref{eqn:rrl_frequency_dependence}.  To account for this, $T_R$ is typically approximated as a fraction of $T_{cont}$.

\begin{figure*}[t]
  \includegraphics[scale=0.7]{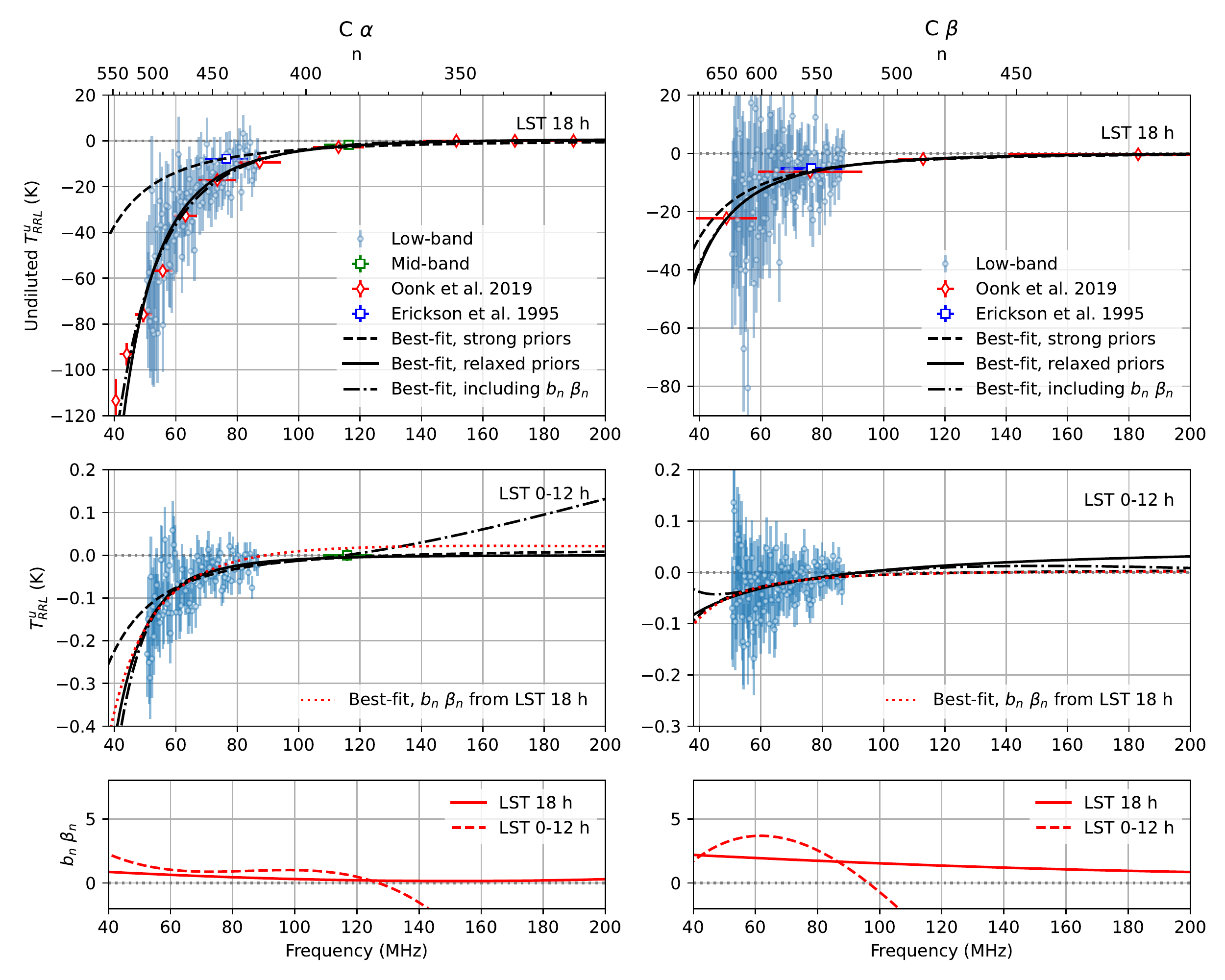}
  \caption{Frequency dependence of C$\alpha$ and C$\beta$ absorption lines.  The top panels shows LST~18~h after correcting for beam dilution and spectral smearing.  The middle panels shows the average over LST~0-12~h assuming the lines are due to diffuse gas filling the beam, hence no beam dilution correction is applied.  In both cases, EDGES low-band data (blue circles) have a strong frequency dependence.  A single mid-band point (green square) for C$\alpha$ represents the average across its analyzed data.  For LST~18~h, the \citet{oonk2019spectroscopy} data (red diamonds) and a composite of the \citet{erickson1995low} data (blue square) are shown after adjusting to brightness temperature units as described in Section~\ref{sec:rrl_frequency_dependence}.  1$\sigma$ error bars are plotted for all points, but do not include dilution uncertainty in the LST~18~h case, which is conservatively estimated as a 50\% error in undiluted $T_{RRL}^{u}$.  Overall, there is good agreement between this work and \citet{oonk2019spectroscopy}.  The black solid and dashed lines show best-fit models in the LTE limit of Equation~\ref{eqn:rrl_frequency_dependence} as described in Section~\ref{sec:rrl_frequency_dependence}.  The black dash-dot lines show the best-fit model when the departure coefficient product, $b_n \beta_n$, is modeled as a 3$^{rd}$-order polynomial in frequency and included in the parameter estimation with the strong priors.  The best-fit departure coefficient models are shown in the bottom panel. For LST~18~h, the best-fit $b_n \beta_n$ for C$\alpha$ qualitatively resembles the curves in \citet{2017ApJ...837..141S}.
  }
  \label{fig:rrl_frequency_dependence}
\end{figure*}

In Figure~\ref{fig:rrl_frequency_dependence}, we show the frequency dependence of the C$\alpha$ and C$\beta$ lines measured by EDGES.  To place the EDGES observations on the same scale as previously reported results from other instruments, we use undiluted $T_{RRL}^{u}$ line amplitude estimates instead of the directly measured values.  
To estimate the undiluted $T_{\text{RRL}}^u$ of individual lines, we use the amplitude of the continuum-subtracted residual spectrum at each line center frequency. The $1\sigma$ uncertainty for each line is estimated from the residual spectrum RMS, scaled to each line frequency by a power-law matching the sky spectrum.  For LST~18~h, we plot the undiluted estimates and compare them with the \citet{oonk2019spectroscopy} and \citet{erickson1995low} observations of the Galactic Center region.  The \citet{oonk2019spectroscopy} data were reported as $\tau d\nu$ integrals, where $\tau=T_{RRL}/T_{cont}$ is the observed line-to-continuum ratio.  To convert them to brightness temperature units, we estimate $\tau$ from their reported $\tau d\nu$ and FWHM$_{line}$ and then multiply by $T_{cont}$ assuming  $\beta=-2.4$ and $T_{100}=10,000$~K, appropriate for the inner Galactic Plane. We use the reported non-detections as $\tau d\nu =0$, and the associated upper limits as error bars on $\tau d\nu$. We also calculate and correct for beam dilution factors for each of their measurements using their reported FWHM$_{beam}$ values and treating them as the diameters of  circular top hat areas in solid angle.  The beam dilution factors increase from 0.63 at 40~MHz to unity (no dilution) at 87~MHz and above. \cite{erickson1995low} report the line amplitudes scaled by the system temperature. With a FWHM$_{beam}$ of 4$\degree$ at 76.4~MHz, the line amplitudes are likely not beam diluted and we apply no correction. We take 35 of the total 43 reported fields from the line-forming region with Galactic longitudes $l=340 \degree$ to $l=20 \degree$, ignoring 8 regions with unreported system temperature and those with non-detections. We rescale them with the reported system temperature and calculate an effective mean line amplitude of $-7.9 \pm0.9$~K at an effective mean frequency of 76.4~MHz for C348$\alpha$-C444$\alpha$. Even with the simple assumptions and approximations used to place all of the observations on the same scale, we see good agreement between the line amplitudes measured here and previously reported by \citet{oonk2019spectroscopy} and \citet{erickson1995low}.  We repeat a similar process for the middle row of Figure~\ref{fig:rrl_frequency_dependence} to show the undiluted $T_{\text{RRL}}^u$ of C$\alpha$ and C$\beta$ lines measured by EDGES away from the inner Galactic Plane in a large LST~0-12~h averaged bin.  In this case, however, there are no comparable observations from other instruments for comparison.

As a further check, we fit a model to the data using Equation~\ref{eqn:rrl_frequency_dependence} simplified to its LTE limit by setting $b_n=\beta_n=1$.  For LST~18~h, we fit all three data sets jointly and initially assume $T_e=100$~K and $T_R(\nu) = 0.2 \times T_{cont}(\nu)$, similar to the background radiation model used by \citet{oonk2019spectroscopy}.  We again use $T_{100}=10,000$~K and $\beta=-2.4$.  We find a best-fit $\tau_L= 1.91 \times 10^{-3}$ for C$\alpha$ with Bayesian information criterion (BIC) of 2938. This is shown as the ``strong prior'' model in Figure~\ref{fig:rrl_frequency_dependence}.  Removing our assumptions for $T_e$, $T_{100}$, and $\beta$, and including them in the parameter estimation, improves agreement between data and model and is highly preferred, decreasing the BIC to 844.  This ``relaxed prior'' fit yields $T_e=190$~K, $T_{100}=5550$~K, $\beta=-3.6$, and $\tau_L= 5.01 \times 10^{-3}$. Similarly, for C$\beta$, a strong prior fit yields $\tau_L= 1.62 \times 10^{-3}$  with a BIC of 899, while a ``relaxed prior'' fit yields $T_e=162$~K, $T_{100}=9000$~K, $\beta=-2.7$, and $\tau_L=1.82 \times 10^{-3}$ with a BIC of 854, thus favoring a relaxed prior model.  While the relaxed priors yield good fits in both cases, the steep spectral index in the best-fit C$\alpha$ model is unrealistic.

For the LST~0-12~h case, which contains only EDGES observations, we test an analogous pair of strong and relaxed prior constraints.  For the strong prior, we use $T_{100}=1000$~K, but retain $T_e=100$~K and $\beta=-2.4$.  For C$\alpha$, we find $\tau_L=1.29 \times 10^{-4}$ with a BIC of 281.  With relaxed priors we find best-fit $T_e=10$~K, $T_{100}=42,965$~K, $\beta=-4.1$, and $\tau_L=1.10 \times 10^{-6}$ with a BIC of 271, thus slightly favoring a relaxed prior model, but hitting the lower limit on the prior set to $T_e$ and finding an extreme $T_{100}$. Similarly for C$\beta$, a strong prior model yields $\tau_L=4.90 \times 10^{-5}$ with a BIC of 366, while a relaxed prior model yields $\tau_L = 2.09 \times 10^{-4}$, $T_e=278$~K, $T_{100}= 1285$~K, $\beta=-1.0$ with a BIC of 353, although here we hit the lower limit on the prior for $\beta$. The relatively low signal to noise ratio of the C$\beta$ observations and more limited frequency range at LST~0-12~h compared to LST~18~h likely reduces the sensitivity to distinguish between the two models for the C$\beta$ lines. Nevertheless, we again see unrealistic background radiation preferred in the relaxed prior cases.

 
For both LST~18 and 0-12~h cases for C$\alpha$, the relaxed prior models prefer steep spectral indices for $T_R$ that are physically unrealistic.  This suggests assuming ideal LTE conditions may be insufficient to accurately capture the full frequency dependence of the line amplitudes along the inner Galactic Plane between 40 and 200~MHz. A full treatment of $b_n$ and $\beta_n$ in the model could improve agreement with data without driving the models to extreme background radiation properties.  It is beyond the scope of this work to incorporate detailed departure coefficient models, which are sensitive to $T_e$ and $n_e$ for each of the gas clouds in the observed fields, but we can test if smoothly varying departure coefficients can provide useful degrees of freedom for the model to be consistent with the data without requiring extreme conditions for the gas or background radiation.  Since $b_n$ has been calculated to remain close to unity and the background radiation brightness temperature is likely much larger than the gas temperature, we expect that RRL amplitudes in our observations are dominated by the $T_R$ term in Equation~\ref{eqn:rrl_frequency_dependence} and relatively insensitive to the $T_e$ term.  We keep $b_n=1$ fixed and focus on the product $b_n \beta_n=\beta_n$, for both C$\alpha$ and C$\beta$ cases. The departure coefficient models of \citet{2017ApJ...837..141S} show that $\beta_n$ is generally a monotonically varying smooth function for $300 \lesssim n \lesssim 550$ for C$\alpha$, and $490 \lesssim n \lesssim 700$ for C$\beta$, over a large range of $T_e$ and $n_e$.  We account for this structure by modeling $\beta_n$ as a 3$^{rd}$-order (cubic) polynomial in frequency as:
\begin{equation}
\label{eq: spectral_dependence_hyd}
\beta_n = \sum^M_{m=0} {\beta_n}_m  \left ( \frac{\nu_n}{\nu_{100}} \right )^m
\end{equation}
with $M=3$.  We fit for the $\beta_n$ terms and $\tau^{LTE}_L$ while otherwise following the strong prior scenario described above with fixed $T_e$, $T_{100}$, $\beta$.  For C$\alpha$ in the LST~18~h case, including the polynomial for $\beta_n$ yields good fits and is the most preferred model, lowering the BIC by 64 beyond the relaxed prior case, whereas for C$\beta$ at LST~18~h, including $\beta_n$ is slightly less preferred increasing the BIC by 5 over the relaxed prior case. For LST~0-12~h, including the $\beta_n$ polynomial is not preferred for both C$\alpha$ and C$\beta$, with similar BIC to the relaxed prior model. Further, for the C$\beta$ model fit, $\beta_{n1}$ and $\beta_{n3}$ hit their prior limits.

\begin{table*}[!htbp]
\label{tab:spectral_dependence}
\centering
\caption{\label{tab:spectral_dependence} Best-fit models of RRLs}

\begin{tabular*}{0.9\textwidth}{c || c cccc || cccccccc}

\hline \hline
Line & LST & \multicolumn{4}{c}{LTE} & \multicolumn{8}{c}{non-LTE}\\
\hline

& & $\tau^{LTE}_L$ & $T_e$ & $T_{100}$ & $\beta$ &  $\tau^{LTE}_L$ & $T_e$ & $T_{100}$ & $\beta$ & ${\beta_n}_0$  & ${\beta_n}_1$ & ${\beta_n}_2$ & ${\beta_n}_3$\\ 

 & & & (K)& (K) &  &  &  (K) & (K) & & & & &\\

\hline
C$\alpha$ & 18~h & ${5.01 \times 10^{-3}}$ & 190 & 5550 & -3.6 & $8.71 \times 10^{-3}$& 100 & 10,000 & -2.4 & 0.30 & -0.60 & 0.56 & 0.03  \\

& 0-12~h & $1.10 \times 10^{-6}$ & 10 & 42,965 & -4.1 & $1.26 \times 10^{-4}$ &  100& 1000&  -2.4 &  1.05
  & -0.33 & -7.82 & -17.89  \\
\hline

C$\beta$ & 18~h & $1.82 \times 10^{-3}$ & 162 & 9000 & -2.7  & $9.75 \times 10^{-4}$ & 100 & 10,000 & -2.4 & 1.53 & -0.94 & 0.25 &  0.008 \\

& 0-12~h & $2.09 \times 10^{-4}$ & 278 & 1285 & -1.0 & $1.26 \times 10^{-5}$ &      100 &      1000 &  -2.4 & 0.71 &
 -20.00 &         -14.61 &  20.00 \\

\hline
H$\alpha$ & 18~h & $5.01 \times 10^{-5}$ & 9925 & 5000 & -3.84 & $8.91 \times 10^{-3}$ & 7000 & 10,000 & -2.4 & 0.65 &  -2.29 & 4.37 &  0.59 \\
\hline

\end{tabular*}
\tablecomments{Best fit parameters for LTE and non-LTE models of the frequency dependence of C$\alpha$, C$\beta$ and H$\alpha$ lines using Equation~\ref{eqn:rrl_frequency_dependence}. For the LTE model, $b_n = \beta_n = 1$.  For the non-LTE model, $T_e$, $T_{100}$, and $\beta$ are fixed matching the strong prior cases described in Section~\ref{sec:rrl_frequency_dependence}, $b_n = 1$, and $\beta_n$ is represented by a 3$^{rd}$-order polynomial in frequency following Equation~\ref{eq: spectral_dependence_hyd}.  The data and models are plotted in Figures~\ref{fig:rrl_frequency_dependence} and~\ref{fig:spectral_dependence_hyd}.}
\end{table*}

The bottom panel of Figure~\ref{fig:rrl_frequency_dependence} shows the derived $b_n\beta_n$ dependence on frequency.  For LST~18~h, it matches qualitatively with \citet[their Figure~8]{2017ApJ...837..141S} and has a smoothly varying trend across the frequency range with $0.79<b_n\beta_n<4.5$.  With its higher values at lower frequencies, this non-LTE extension absorbs the need for a steep spectral index in the background radiation found in the LTE limit.  The LST~0-12~h observations are more limited without accompanying measurements at high frequencies and the best-fit is not constrained by data above 120~MHz.  This makes $b_n \beta_n$ highly covariant with $\tau^{LTE}_L$.  We had to fix $\tau^{LTE}_L=1\times10^{-4}$ to prevent the estimation from trending to unrealistically large $\tau^{LTE}_L\approx1$ and small $b_n \beta_n$.  With this additional constraint, we found $0<b_n\beta_n<2.3$ over the range of the data.  If the gas in both LST cases has similar properties, we might expect that the better constrained $b_n \beta_n$ model from the LST~18~h data would also improve the LST~0-12~h fits.  Using that polynomial instead of $b_n=\beta_n=1$, and fixing all other parameters except $\tau^{LTE}_L$ with the strong prior assumptions, decreases the BIC by nearly three, making it the most preferred model for the LST~0-12~h case.

\begin{figure}
    \centering
    \includegraphics[width = \columnwidth]{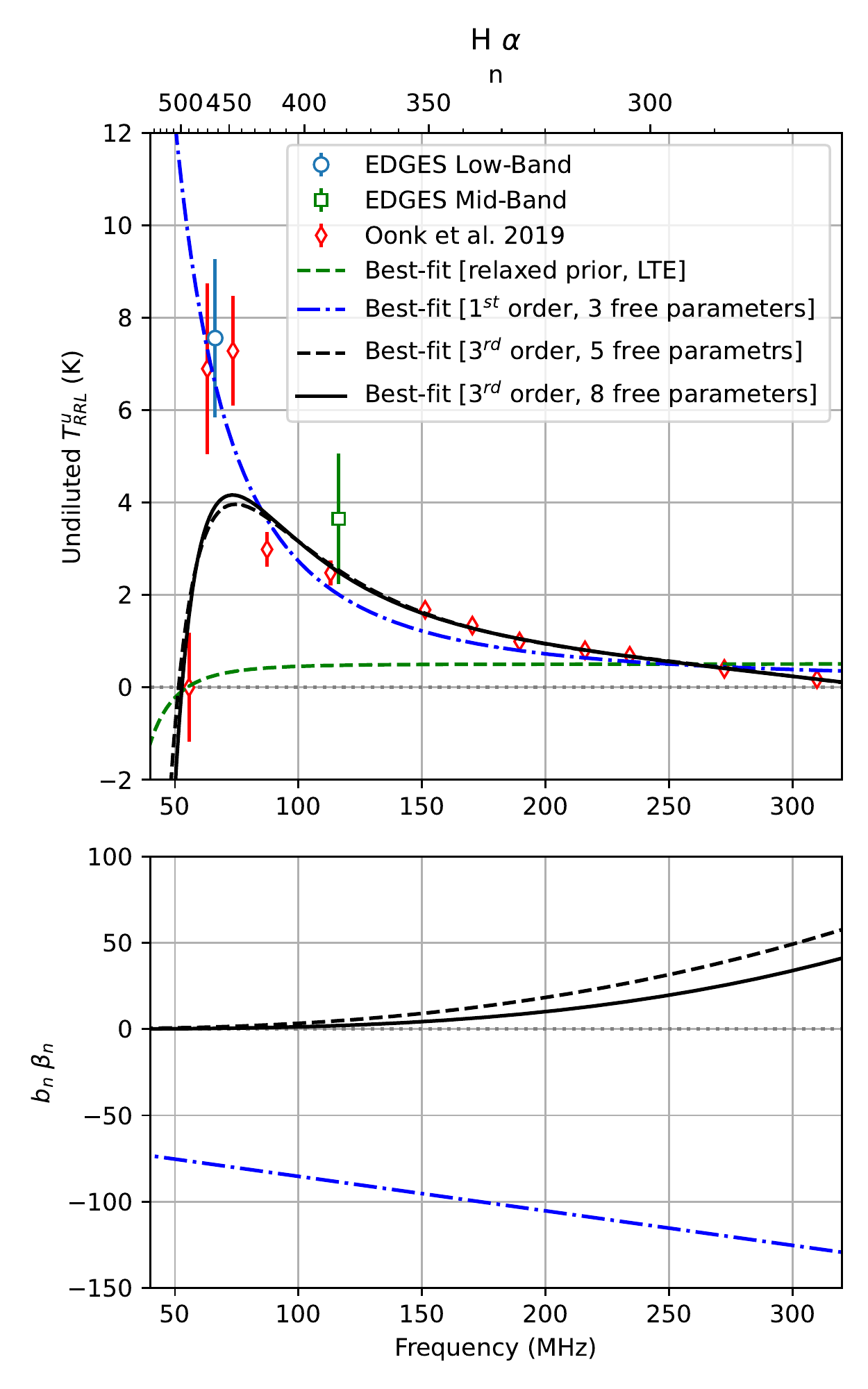}
    \caption{Frequency dependence of H$\alpha$ emisssion lines for the LST~18~h case. We show the undiluted line amplitude of EDGES low-band and EDGES mid-band, along with the results from \citep{oonk2019spectroscopy} after adjusting to brightness temperature units following the same process as for C$\alpha$ and C$\beta$ in Figure~\ref{fig:rrl_frequency_dependence}. Again we see reasonable agreement between the EDGES and \citep{oonk2019spectroscopy} observations. The green dashed line shows the best-fit model in the LTE limit.  The blue dashed line shows the best-fit model for a non-LTE case using the strong priors described in Section~\ref{sec:rrl_frequency_dependence} and with $b_n=1$ and $b_n\beta_n$ modeled as a 1$^{st}$-order (linear) polynomial.  The black solid and dashed lines show the best-fit models in the non-LTE limit of Equation~\ref{eqn:rrl_frequency_dependence} where $\beta_n$ is further expanded as a 3$^{rd}$-order polynomial. The bottom plot shows the derived models of $b_n\beta_n$ in the non-LTE cases.}
    \label{fig:spectral_dependence_hyd}
\end{figure}

We repeat a similar analysis to show the frequency dependence of H$\alpha$ for the LST~18~h case. We apply the beam dilution and spectral smearing factors for EDGES low-band and EDGES mid-band as described in Section \ref{sec:innerplane} and use the H$\alpha$ observations and instrument properties reported in \cite{oonk2019spectroscopy} to place their data on the same temperature scale. Similar to C$\alpha$ and C$\beta$, we see a good agreement between the measurements reported here and those in \cite{oonk2019spectroscopy}, providing further evidence that our simple approximations for placing all observations on the same scale are reasonable. 
We again use Equation~\ref{eqn:rrl_frequency_dependence} and begin by fitting in the LTE limit with relaxed priors. This yields $\tau^{LTE}_L = 5.01 \times 10^{-5}$, $T_e = 9925$~K, $T_{100}=5000$~K, $\beta = -3.84$ with a BIC of 571 and is shown as the best-fit relaxed prior case in Figure~\ref{fig:spectral_dependence_hyd}.  The model is in poor agreement with the data, failing to capture the increase in amplitude with decreasing frequency and again resulting in an unrealisticly steep best-fit spectral index for the background radiation. 

As with the carbon lines, the poor agreement motivates exploring the models for H$\alpha$ in non-LTE conditions.  
To do so, we assume a hot hydrogen gas scenario and use the strong priors to fix $T_e = 7000$~K, $T_{100} = 10,000$~K, $\beta = -2.4$, and $b_n=1$.  We fit only for $\tau^{LTE}_L$ and the $\beta_n$ components of a 1$^{st}$-order (linear) polynomial using $M=1$ in Equation~\ref{eq: spectral_dependence_hyd}. This yields a best-fit with a negative $b_n \beta_n$ component, $\tau^{LTE}_L = 1.58 \times 10^{-5}$, and a BIC of 171. 
The model is now in reasonable agreement with the data, although the lowest frequency measurement of \cite{oonk2019spectroscopy} at 55.7~MHz suggests an unmodeled turnover may occur.  

Extending the $\beta_n$ model to the full 3$^{rd}$-order polynomial ($M=3$) allows it to capture the apparent turnover at the lowest frequencies.  We show the result of using the 3$^{rd}$-order polynomial and the same strong prior constraints as above in Figure~\ref{fig:spectral_dependence_hyd} as the case with five free parameters.  It yields a positive $b_n\beta_n$ contribution with $\tau^{LTE}_L = 8.91 \times 10^{-3}$ and a BIC of 26. Relaxing all priors (labeled as eight free parameters in Figure~\ref{fig:spectral_dependence_hyd}), we find a best-fit $b_n\beta_n$ contribution similar to the strong prior case along with $\tau^{LTE}_L = 0.1$, $T_e = 3954$~K, $T_{100}= 15,400$~K, $\beta = -2.98$ and a BIC of 32. Thus, the non-LTE strong prior case with five free parameters is favored among all models.  Given the sensitivity of $b_n\beta_n$ to the lowest frequency data point reported by \cite{oonk2019spectroscopy}, more observations around and below 50~MHz would be valuable.  

Table~\ref{tab:spectral_dependence} lists the best-fit model parameters for C$\alpha$, C$\beta$, and H$\alpha$ lines in the LTE limit with relaxed priors and in the non-LTE scenario with strong priors and a 3$^{rd}$-order polynomial for $\beta_n$. Note that we only report the best-fit model parameters without their associated statistical uncertainties because errors are likely driven more by biases in our simple assumptions for scaling measurements onto the same temperature scale than by the statistical uncertainties in the fits.

The LTE models presented in \cite{shaver1975theoretical} and \cite{salgado2017low} show increasing amplitudes of H$\alpha$ lines with increasing frequency that are inconsistent with the amplitude changes observed by EDGES from 66~MHz to 116~MHz and with our interpretation of the data reported in \cite{oonk2019spectroscopy}. However including the departure coefficients in the models improves agreement with the data suggesting the more sophisticated non-LTE treatment is necessary to model the spectral dependence of RRLs at the observed frequencies.  Thus, collectively, these findings are consistent with theoretical analyses that show non-LTE effects are significant at large-$n$ for lower density, higher temperature gas, such as \citet[their Figure~5]{shaver1975theoretical}.  Both negative and positive best-fit ranges of $b_n\beta_n$ 
are allowed in the models of \cite{salgado2017low}, depending on the electron temperature and density of the gas.    

We briefly consider alternative explanations for the decrease in H$\alpha$ amplitude with frequency other than non-LTE physics.  Strong H$\alpha$ emission lines may be perferentially flagged during the RFI excision in our analysis, particularly in the mid-band data, however we find no bias in the flag counts at the frequencies corresponding to H$\alpha$ lines. Overlap between the H$\alpha$ and C$\alpha$ lines could hinder the model fitting, however our synthetic tests on two-component model with varying separation of C$\alpha$ and H$\alpha$ lines (20~kHz-100~kHz) shows confident retrieval of line parameters with less than 5\% error for all line parameters that have at least 20~kHz seperation, while the model fails to retrieve the hydrogen line parameters at 10 kHz separation, and the errors on carbon line parameters rise to 15\%. This establishes robsutness against such confusion unless there were an unknown systematic offset between the radial velocities of hydrogen and carbon clouds that causes them to overlap more than expected.  


\cite{oonk2019spectroscopy} report evidence for a nearby cold hydrogen cloud along the line of sight to the Galactic Center based on a narrowing of the H$\alpha$ line width with decreasing frequency from 25~km/s between 300~and 80~MHz to 12~km/s in their lowest frequency detections at 73~and 63~MHz.  This could be explained by beam dilution effects as the beam size increases at lower frequencies it could encompass more of an extended nearby cloud of cooler gas compared to the Galactic Center region.  Such effects are less likely for the EDGES observations, which have nearly constant beam size, hence the relative dilution of different regions should not change with frequency. The other possible explanations could be due to line blending with C$\alpha$ lines, however we see a similar scaling in line width as that of \cite{oonk2019spectroscopy} (a factor of 2.5 from $12 \pm 3.7$~km/s at 63~MHz and $29 \pm 3.7$~km/s at 113~MHz for \cite{oonk2019spectroscopy}, and a factor of 2.8 from $77 \pm $9~km/s at 66.7~MHz and $221.6 \pm 68$~km/s at 116 MHz for EDGES). Further, we find the \cite{oonk2019spectroscopy} observations are consistent with the $T_{RRL}^{u}$ derived from EDGES, suggesting little change in line blending due to frequency-dependent beam dilution since the two instruments have different beam characteristics---although the uncertainties are large in both cases, suggesting the need for more precise measurements.  

\section{Implications for 21~cm Observations}
\label{sec:21cm}
  Summarizing the findings from Section~\ref{sec:3.3} for diffuse RRL contributions away from the Galactic Center, we have detected $C\alpha$ absorption at about 30-40~mK and placed upper limits on C$\beta$ absorption and H$\alpha$ emission lines at 30~mK levels in the 50-87~MHz band.  In the 108-124~MHz band, we have placed upper limits on C$\alpha$ absorption and H$\alpha$ emission at about 14~mK.  These levels are generally at or below the expected redshifted 21~cm angular and spectral fluctuation amplitudes in theoretical models (e.g., \citealt{furlanetto2004observing, barkana2009studying, 2019MNRAS.486.1763F}), which are typically 10-100~mK, depending on model and redshift. Here we empirically study the quantitative effects induced by RRLs on the global and power spectrum of 21~cm signal using the amplitudes of the detected lines. 



\subsection{Global 21~cm Signal}
\label{sec:G21_effects}
Early results in the search for the global 21~cm signal (e.g. \citealt{monsalve2017results, bowman2018absorption, 2018ApJ...858...54S, singh2022detection}) have generally ignored the presence of RRLs.  To test if this is a reasonable assumption, we create simulated observations to investigate the recovery of the global 21~cm signal with and without the strongest RRLs in an extreme scenario assuming the measurement is made at LST~18 when the strong lines on the inner Galactic Plane are present.  

To model the carbon absorption lines, 
we include all individual C$\alpha$ lines in the full EDGES low-band of 50-100~MHz by scaling the measured stacked line amplitude in frequency and adding Gaussian profiles with 19.8~kHz FWHM at each line center.  For the frequency scaling, we note that for the target frequencies, Equation~\ref{eqn:rrl_frequency_dependence} is dominating by the $T_R$ term.  We therefore simplify the expression and use only a power-law:
\begin{equation}
T_{RRL}(\nu_n) \approx A \left (\frac{\nu}{\nu_{eff}} \right )^\beta  
\end{equation}
where $A$ is the best-fit stacked amplitude at LST~18~h from Table~\ref{tab:fitparamsalpha}, $\nu_{eff}=66.1$~MHz is the effective frequency of our observations, and $\beta$ is the best-fit spectral index for the relaxed prior LTE model in Table~\ref{tab:spectral_dependence}.  For C$\alpha$ lines we use $A=-0.795$~K and $\beta=-3.6$, while for C$\beta$ we use $A=-0.261$~K and $\beta=-2.7$.

H$\alpha$ emission lines are added with a constant amplitude of 203~mK, as an extreme case of the highest detected amplitude representing all the lines in the full frequency band.  This RRL model has an RMS of 207~mK for a simulated spectrum with 6~kHz resolution, matching the raw resolution of the EDGES data.  The RMS reduces to 101~mK after binning to 61~kHz, which matches the spectral resolution of SARAS \citep{singh2022detection}, and 14~mK after binning to the 390~kHz resolution used in \cite{bowman2018absorption}. The nosie falls faster than the square root of the channel size because, for large spectral bins, some of the emission and absorption lines fall in the same bins, offsetting their contributions to the RMS.  At the respective spectral resolutions, the RMS is below the thermal noise, which is about 25~mK for deep EDGES observations and about 200~mK for SARAS \citep{singh2022detection}.  Hence we expect the RRLs will have little impact on the signal recovery. 

\begin{figure}[t]
  \centering
  \includegraphics[scale=0.55]{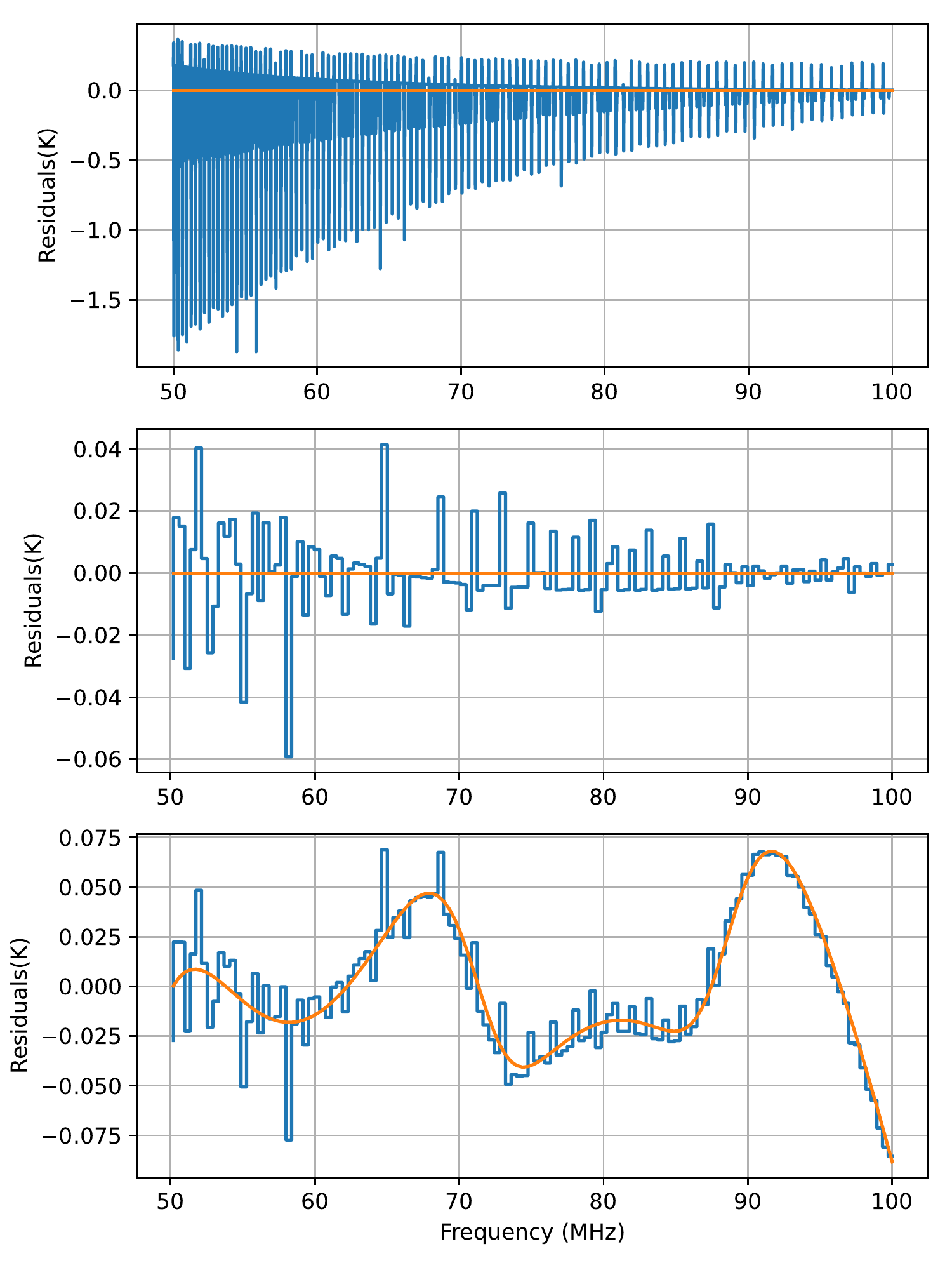}
  \caption{  \textit{Top:}  Residuals after fitting and removing a 5-term foreground model from simulated spectra in the 50-100~MHz band at LST~18~h with 6~kHz resolution.  The orange curve shows the residuals to a model without RRLs, whereas the blue curve includes modeled C$\alpha$, C$\beta$, and H$\alpha$ lines. \textit{Middle:} The residuals rebinned to 390~kHz for comparison with \citet{bowman2018absorption}. \textit{Bottom:} Residuals after the foreground model is fit and removed when a 200~mK flattened Gaussian global 21~cm absorption signal is also included in the simulated spectra.  RRLs contribute to an increase in RMS by 12~mK in the data.  The difference in amplitude of the best-fit 21~cm profile is only 2~mK, an error of 1\%.}
  \label{fig:global21cm}
\end{figure}

To test signal recovery, we start by extracting a realistic foreground continuum model from the EDGES observations used in \cite{bowman2018absorption}. We use \texttt{edges-analysis}\footnote{\href{https://github.com/edges-collab/edges-analysis}{https://github.com/edges-collab/edges-analysis}} to calibrate the data and flag RFI, correct for beam chromaticity, and bin the observations between LST 17-18~h.  We simultaneously fit a 5-term linearized logarithmic polynomial (``linlog'') foreground model and a 5-term flattened Gaussian 21~cm absorption profile.  We use this best-fit foreground model as the continuum in the simulated observations.  We then add the RRL model and a modeled 21~cm absorption feature matching the best-fit properties reported in \cite{bowman2018absorption}, but with a conservative amplitude of 200~mK. When we fit a 5-term linlog foreground model and 21~cm absorption flattened Gaussian profile  simultaneously to the simulated observations, the difference in amplitude of the best-fit global 21~cm absorption feature with and without the RRLs is about 2~mK, extremely small compared to the 200~mK injected signal.  Figure~\ref{fig:global21cm} shows the RRL model and resulting global 21~cm fit.  Lastly, we tested flagging RRLs in the analysis of actual observations and found it did not affect the results reported in \cite{bowman2018absorption}.

These findings apply directly to EDGES observations.  They also should be representative of most other global 21~cm experiments since most instruments have large beams that would similarly dilute the observed amplitude of strong lines from the inner the Galactic Plane.  We note, however, the effects of RRLs on global 21~cm searches could be more significant for instruments with higher angular resolution.

\begin{figure}[t]
  \centering
  \includegraphics[scale=0.55]{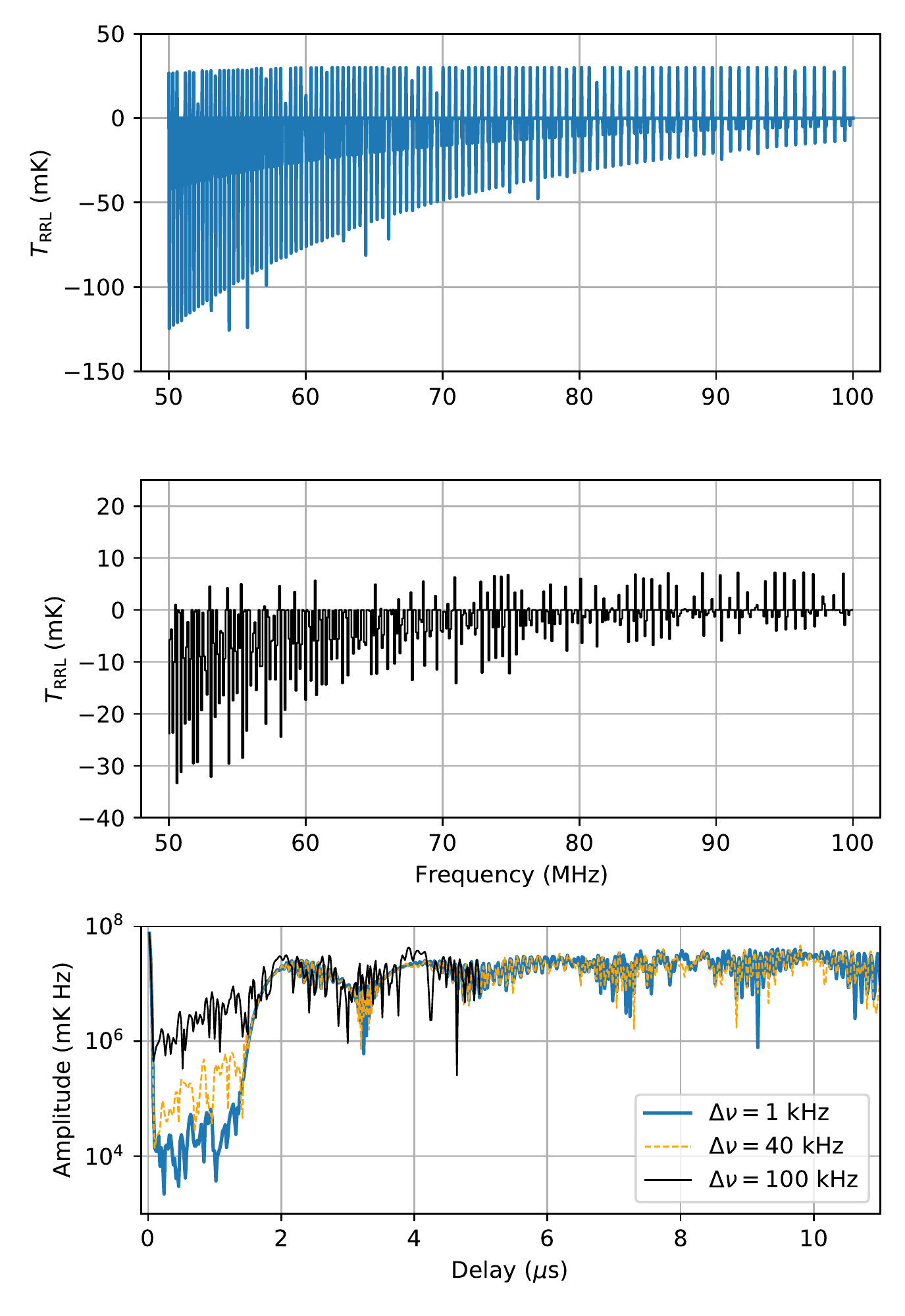}
  \caption{ \textit{Top:}  Simulated RRL spectrum, similar to Figure~\ref{fig:global21cm} but for LST~0-12~h, away from the inner Galactic Plane.  C$\alpha$, C$\beta$ and H$\alpha$ lines are modeled in the 50-100~MHz band with 1~kHz resolution.  No other foreground contributions are included.  \textit{Middle:} The simulated spectrum rebinned to 100~kHz, which is approximately the spectral resolution used by HERA. \textit{Bottom:} Delay spectra calculated from the simulated RRLs for three different spectral binnings.  These are equivalent to delay spectra from a notional zero-length baseline in an interferometer.  Most of the power contributed by the RRLs is above 2~$\mu$s delay, corresponding to spectral scales smaller than 500~kHz.  The 100~kHz spectral resolution of HERA couples power from smaller scales to larger scales of order 1~MHz.  An intermediate spectral resolution of 40~kHz removes much of the coupling to large scales. Even with coupling, the power from scales larger than 500~kHz ($< 2$~$\mu$s) contributes less than 10~mk$^2$ to the total 36~mk$^2$ variance in the simulated spectrum with HERA-like resolution. }
  \label{fig:delay_lst0-12}
\end{figure}

\subsection{21~cm Power Spectrum}
\label{sec:21cm_powerspectrum}

Interferometers optimized for 21~cm power spectrum observations aim to detect and characterize the IGM during Cosmic Dawn and reionization.  Most instruments have focused on observing the reionization era between $13>z>6$, corresponding to frequencies between 100~and 200~MHz.  Recently, several instruments have begun to place initial upper limits on the 21~cm signal during Cosmic Dawn between $25>z>15$, between about 50~and 90~MHz \citep{eastwood201921, gehlot2019first, gehlot2020aartfaac}.   During reionization, fluctuations in the 21~cm signal are dominated by the growth of reionized regions on scales of order 10~Mpc, corresponding to spectral fluctuations on scales of order 1~MHz in observations, although there is also a significant power on smaller scales.  Before reionziation, the 21~cm power spectrum traces the matter power spectrum and ultraviolet and X-ray heating of the IGM, yielding power on generally similar scales as during reionization. 
The amplitude of 21~cm fluctuations is predicted to be $\Delta^2 \lesssim 10^2$~mK$^2$, varying with redshift as the IGM evolves. 
Although there have been significant improvements in upper limits of 21 cm power spectrum measurements in the past decade, the current best upper limits are an order of magnitude higher than the fiducial model \citep{barry2021ska, abdurashidova2022first}. 

Here, we estimate the approximate power contributed by RRLs over these spectral scales in a typical observation by simulating all C$\alpha$, C$\beta$, and H$\alpha$ RRLs spanning 50-100~MHz following a similar method as for the global 21~cm test but using the average line amplitudes and limits found for LSTs 0-12~h.  These lines are much weaker than those along the inner Galactic Plane.  This analysis is directly relevant for the 21~cm interferometers targeting Cosmic Dawn between $25>z>15$, including OVRO-LWA, LOFAR, and AARTFAAC. The Hydrogen Epoch of Reionization Array (HERA; \citealt{deboer2017hydrogen,abdurashidova2022first}) is also sensitive down to 50~MHz, although it is optimized for 100-200~MHz to observe reionization.  We focus our analysis here on HERA given our familiarity with the instrument.  

To simulate observations including RRLs, we use the best-fit strong prior model shown in the middle panel of Figure~\ref{fig:rrl_frequency_dependence} as our reference for C$\alpha$ amplitudes and a constant H$\alpha$ amplitude of 33~mK across the band, which is the average from the individual two-hour LST bins centered between 0-12~h.  This model somewhat overestimates the line strength compared to the very weakest region of sky at LSTs 2-6~h, but will be more typical of the region covered by long drift scans with HERA. We note that instruments like HERA have high angular resolution (0.8$\degree$ at 75~MHz) in comparison to EDGES (72$\degree$ at 75~MHz).  Here we study a monopole-like RRL contribution that is assumed to be uniform over the full beam of HERA, focusing on the spectral line-of-sight effects rather than the angular effects. The resulting simulated spectra have 14~mK RMS when the lines are fully resolved at 1~kHz resolution, falling to 6~mK RMS when binned with 100~kHz resolution, typical of HERA observations. We note that due to the smaller beam of HERA, one could expect lower line blending as compared to EDGES.

Figure~\ref{fig:delay_lst0-12} shows the simulated RRL contribution to HERA spectra and the corresponding delay spectra for different spectral resolutions.  The delay spectra are found by applying a Blackman-Harris window function to the simulated sky spectra and Fourier transforming into delay space.  The delay spectra are equivalent to what would be seen by a zero-length baseline in an interferometer if the sky contained only RRLs.  They illustrate that much of the power introduced by the RRLs resides on small spectral scales below 500~kHz, corresponding to delays greater than 2~$\mu$s.  The limited spectral resolution of HERA couples some of this small-scale power to larger scales, but the total power contributed by scales larger than 500~kHz accounts for less than 10~mK$^2$ of the total 36~mK$^2$ variance in the 100~kHz-binned spectrum. These variances are two orders of magnitude lower than the current best upper limits at high redshift of $\Delta^2 \leqslant 3496$~mK$^2$ at $z=10.4$ \citep{abdurashidova2022improved} and $\Delta^2 \leqslant 7388$~mK$^2$ at $z=17.9-18.6$ \citep{gehlot2020aartfaac}.  The RRL power will likely be lower on non-zero length baselines, falling quickly with baseline length similar to other diffuse Galactic emission. Future work to map the spatial distribution of RRLs would be valuable to confirm this assumption.  The levels of systematics introduced due to RRLs may need to be considered in next generation intruments.  Nevertheless, the power levels introduced by RRLs are at the low end of expected 21~cm signal levels and we anticipate the lines can likely be ignored in the current generation of 21~cm interferometers, which are focused on detection and not detailed characterization of the signal.  

In the ultimate case that the lines cannot be neglected initially, two scenarios are possible.  First, if our assumption that diffuse weak lines dominate the observations is incorrect and instead strong lines from several small regions do indeed dominate, then the diffuse contribution would be even lower than our estimated upper limits above.  Surveys of the localized dominant regions would likely enable them to be subtracted from 21~cm observations during processing with reasonable accuracy.  Second, if diffuse lines are present across large areas of the sky at levels that interfere with the 21~cm analysis as may be possible (see e.g., \citealt{grenier2005clouds}), then those frequencies may need to be completely omitted from analysis.   

\begin{figure}[t]
  \centering
  \includegraphics[scale=0.6]{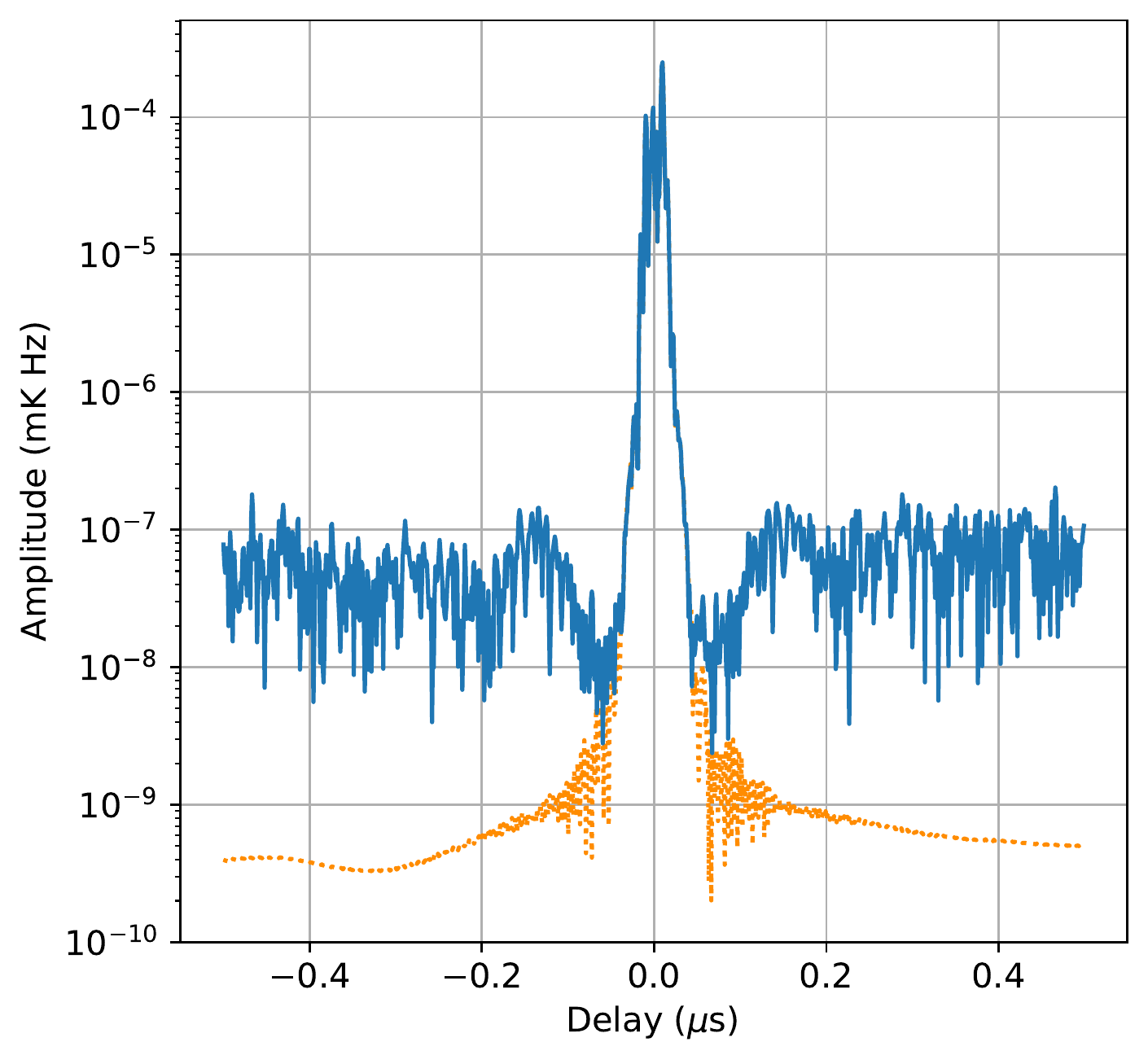}
  \caption{Comparison of simulated delay spectra for a baseline on the HERA radio telescope at LST~6~h  with and without assuming RRL frequencies need to be masked.  The dotted line shows the reference delay spectrum without flagging RRLs. Masking and inpainting RRL frequencies before the delay transform (solid line) raises the foreground leakage by over two orders of magnitude in the delay modes used for redshifted 21~cm science.  A similar trend was observed for the other simulated LSTs and baselines in HERA.}
  \label{fig:fig15}
\end{figure}

To investigate possible spectral leakage effects that RRL flagging could have on 21~cm power spectrum analysis, we use \texttt{pyuvsim} to simulate observations with HERA. HERA is a 350~elements radio interferometer being built in South Africa for observing redshifted 21~cm power spectrum. HERA is designed to observe in the frequency range of 50-250~MHz. We simulate HERA visibilities at LST~0, 6, 12, and~18~h  on 18~baselines for the telescope's east-west polarization using the Global Sky Model \citep{de2010global}. The resulting visibilities are then passed through a Blackman-Harris window and transformed into delay space both before and after masking RRL channels. We apply a delay-inpainting algorithm in the masked delay spectra \citep{parsons2009,aguirre2022validation} and attempt to reconstruct the ideal observation as if RRLs had not been present.  The inpainting algorithm is a deconvolution that effectively extrapolates along frequency for each visibility to fill in gaps due to RFI excision or other missing data.  An example of the resulting delay spectra is shown in Figure~\ref{fig:fig15} for LST~6~h, which is broadly representative of all baselines and LSTs simulated.  The effect of masking the RRL frequencies, even with inpainting, is to raise the noise floor in the modes of interest to 21~cm power spectrum analysis with absolute delays greater than about 0.1~$\mu$s by about two orders of magnitude.  This limits the dynamic range to approximately $10^3$, below the target performance of the telescope that is likely needed to detect the 21~cm power spectrum.  This suggests further development would be needed to identify successful strategies for mitigating diffuse RRLs in power spectrum analysis if flagging is required. 
\\

\section{Conclusions}
\label{sec:4}
We have used existing long drift scans from the EDGES low-band and mid-band instruments, with an effective spectral resolution of 12.2~kHz, divided into LST bins, to assess the strength of RRLs averaged over large areas of the sky. In the lower frequency range of 50-87~MHz, C$\alpha$ absorption lines were seen at all LST periods, with amplitudes from -33~to -795~mK and H$\alpha$ lines were seen in emission, with average fitted amplitudes as a function of LST ranging from no detection to 203~mK. C$\beta$ and C$\gamma$ lines were also seen in absorption with average fitted amplitudes between no detection and -305~mK, and between no detection and $-171$~mK, respectively. At the higher frequency band of 108-124.5~MHz, RRL lines were seen only when the Galactic Center was overhead.  H$\alpha$ lines were seen in emission with amplitude of 98~mK and C$\alpha$ in absorption with an amplitude of -46~mK. We note that due to large beam and line broadening, some line blending of absorption and emission lines and may not be fully captured in our models.

Conservatively interpreting the observations away from the Galactic Center as RRLs from diffuse gas, as opposed to isolated sources, we find that the amplitudes of the diffuse lines are at or below the expected redshifted 21~cm signal from Cosmic Dawn. They will likley have no impact on global 21~cm detection (assuming the global 21~cm signal matches the current fiducial theoretical expectations of 50-100~mK) and are two orders of magnitude below current 21~cm power spectrum limits at high redshift. This suggests Galactic RRLs do not impose significant systematics in low frequency regimes and do not need mitigations in current 21~cm experiments. However, the power levels introduced by RRLs may need mitigation in the next generation of instruments aiming to characterize the 21~cm signal in detail after initial detections.  A delay spectrum analysis for simulated HERA observations showed that if masking and inpainting RRL frequencies is required, foreground leakage could increase by about two orders of magnitude in the higher delay modes used for 21~cm analysis.  This reduces the dynamic range in delay spectra to about $10^3$, suggesting new development would be needed in such a scenario.  Further work is also warranted to extend this analyis to higher frequencies between 100~and 200~MHz, corresponding to the 21~cm redshifts for reionziation, although the non-detection of C$\alpha$ and H$\alpha$ we find away from the inner Galactic Plane at 116~MHz suggests RRLs will be less of a problem for reionization observations than for higher redshift Cosmic Dawn observations.

\begin{acknowledgements}

This work was supported by the NSF through research awards for EDGES (AST-1609450, AST-1813850, and AST-1908933) and by the NASA Solar System Exploration Research Virtual Institute cooperative agreement number 80ARC017M0006. The authors thank the anonymous reviewer for detailed inputs and suggestions. The authors also thank Dr. Anish Roshi for useful discussions and providing relevant references. EDGES is located at the Inyarrimanha Ilgari Bundara, the CSIRO Murchison Radio-astronomy Observatory. We acknowledge the Wajarri Yamatji people as the traditional owners of the Observatory site. We thank CSIRO for providing site infrastructure and support.
\end{acknowledgements}

\software{astropy \citep{2013A&A...558A..33A,2018AJ....156..123A},  
          edges-analysis (\url{http://github.com/edges-collab}),
          pyuvsim \citep{2019JOSS....4.1234L}, pyuvdata \citep{pyuvdata2017}
          }


\bibliography{references}{}
\bibliographystyle{aasjournal}

\end{document}